\documentclass[aps, prd, preprint, preprintnumbers, showpacs, showkeys, amsfonts, nofootinbib, floatfix,11]{revtex4-1}
\usepackage{graphicx}
\usepackage{epsfig}
\usepackage{rotate}
\usepackage{rotating}
\usepackage{multirow}
\usepackage[T1]{fontenc}
\usepackage{array}
\usepackage{graphicx,
}
\usepackage{epstopdf}
\usepackage{amsmath,amssymb,hyperref,enumerate}
\usepackage{pstricks}
\usepackage{slashed}
\usepackage{url}
\usepackage{color}
\hypersetup{colorlinks,citecolor= nicegreen,linkcolor= nicered}
\definecolor{nicered}{rgb}{0.7,0.1,0.1}
\definecolor{nicegreen}{rgb}{0.1,0.5,0.1}

\usepackage[normalem]{ulem}

\def\({\left(}
\def\){\right)}
\def\[{\left[}
\def\]{\right]}

\newcommand{\simgt}{\lower.7ex\hbox{$\;\stackrel{\textstyle>}{\sim}\;$}}
\newcommand{\simlt}{\lower.7ex\hbox{$\;\stackrel{\textstyle<}{\sim}\;$}}

\begin{document}
\renewcommand{\thefootnote}{\fnsymbol{footnote}}
\vspace{-0.5cm}
\preprint {RECAPP-HRI-2013-022; CUMQ/HEP 177}
%hep-ph/
\vspace{1.2cm}

%\vspace*{-0.5cm}

\title {Higgs data confronts Sequential Fourth Generation Fermions in the Higgs Triplet Model}

%\vspace{0.5cm}

\author {Shankha Banerjee$^{(1)}$\footnote{Email: shankha@hri.res.in}}
\author{Mariana Frank$^{(2)}$\footnote{Email: mfrank@alcor.concordia.ca}}
\author{Santosh Kumar Rai$^{(1)}$\footnote{Email: skrai@hri.res.in}}
\affiliation{$^{(1)}$Regional Centre for Accelerator-based Particle Physics, Harish-Chandra Research Institute,  Chhatnag Road, Jhusi, Allahabad 211019, India, }
\affiliation{$^{(2)}$Department of Physics, Concordia University, 7141
Sherbrooke St. West, Montreal, Quebec, Canada H4B 1R6}

%\date{\today}

\begin{abstract}
We investigate the effect of introducing a sequential generation of chiral fermions in the Higgs Triplet Model with nontrivial mixing between the doublet and triplet Higgs.  We use the available Large Hadron Collider data for Higgs boson production and decay rates,  the constraints on the fourth generation masses, and impose electroweak precision constraints from the S, T and U parameters. Our analysis shows that  an SM-like Higgs boson state at $\sim 125$ GeV can be accommodated in the Higgs Triplet Model with four generations, and  thus,  that four generations survive collider {\it and} electroweak precision constraints in models beyond SM. 

\end{abstract}

%%%%%%%%%%%%%%
%%%%%%%%%%%%%%
%\vspace*{-0.9cm}
\maketitle
%%%%%%%%%%%%%%

%%%%%%%%%%%%%%%%%%%%%%%%%%%%%%%%%%%%%%%%%%%%%%%%%%%%%%%%%%%%%%%
\section{Introduction}
The new Large Hadron Collider (LHC) data from ATLAS \cite{ATLAS} and CMS \cite{CMS} seems to indicate a Higgs boson that 
may be consistent with the Standard Model (SM) one. 
%but with questions about the di-photon and $\tau^+\tau^-$ decay branching ratio still remain. 
Although the statistics are not very strong yet, and more analyses are needed, the question on whether the SM is the final theory still remains. In particular, the SM fails to explain observed experimental phenomena such as neutrino masses, dark matter, or  baryogenesis. One could reasonably ask if the Higgs boson found at the LHC, even if finally shown to be consistent with the SM predictions, could not possibly belong to a more complete theoretical scenario,  where the neutral Higgs signatures are consistent with the SM ones.  This situation is akin to requiring new physics scenarios to satisfy low energy precision measurements.

To test this hypothesis, we apply it to one of the simple generalizations of the SM, the Higgs Triplet 
Model (HTM), to which a sequential generation of chiral fermions is added. The addition of a fourth sequential generation of fermion doublets is a natural extension of the SM (SM4) \cite{holdom}. The model restricts fourth-generation quark masses to be not too large to preserve perturbativity \cite{marciano}, and it  does not conflict with electroweak precision observables
\cite{holdom}, as long as their mass differences are small
\cite{yanir}. Further limits on the fourth generation fermion masses exist from direct 
searches at collider experiments such as the LEP, Tevatron, as well as from the current LHC data. 

There are many advantages of introducing an extra family of fermions:  
\begin{itemize}
\item{The fermions associated with the fourth generation could trigger dynamical electroweak symmetry
  breaking \cite{marciano}  without a Higgs boson, and thus deal with 
  the hierarchy problem.} 
\item{ Gauge couplings unification can in principle be achieved without supersymmetry \cite{hung}.}
\item{ A new family might resolve SM problems in flavor physics, such as the CP-violation in $B_s$-mixing \cite{soni}. While the electroweak precision data constrains the mass splitting between the fourth 
generation quarks, data from B--meson physics constrain their mixing pattern \cite{Constraints}.}
\item {A fourth generation could solve problems related to baryogenesis because an additional quark 
doublet could contribute to an increase in the amount of of CP-violation \cite{hou}. }
\item {A fourth generation extension of the SM would increase the strength of the
  electroweak phase transition  \cite{carena}.} 
\item {A fourth generation neutrino can serve as a candidate for cold dark matter \cite{Lee:2011jk}, resolving this outstanding problem of the SM.}
\item{New heavy fermions lead to new interesting effects due to their large Yukawa
couplings \cite{hung2}. }
\end {itemize}

However, the SM4 scenario is severely constrained by the available data \cite{Constraints}. First, 
from constraints on the invisible width of the $Z$ boson at LEP,  the number of
light neutrinos is  $N_\nu=2.9840 \pm 0.0084$  \cite{particledata} and thus the fourth-family neutrino 
must be heavier than $M_Z/2$, assuming
 small mixing with the lighter SM leptons.  A heavy charged lepton with a mass   $m_{\ell'}\!
<\! 100$ GeV has also been excluded at LEP2 \cite{particledata}.  The
Tevatron and now the LHC  have excluded light fourth generation
quarks. Direct searches have been performed by both the ATLAS and CMS
Collaborations, with the CMS Collaboration putting the strongest bound  on the masses of 
degenerate fourth generation quarks, ruling out $m_{q'} < 685$ GeV  at 95\% C.L. \cite{Chatrchyan:2012fp}\footnote {In these experimental
bounds, assumptions such as BR$(b' \to tW)$ or BR$(t' \to bW)=1$ are made;
relaxing them leads to slightly weaker bounds as discussed in e.g. \cite{SM4relaxed}. Further softening 
of the constraints happen for non-degenerate choice of masses for the up-type and down-type quarks of 
the fourth generation.}. Updated bounds can also be extracted from an inclusive search done 
by the CMS Collaboration for vector-like top partners at $\sqrt{s}=8$ TeV \cite{Chatrchyan:2013uxa}. 
Unitarity requirements indicate that fourth generation fermions
should not be extremely heavy, $m_{q'} < 500$ GeV  \cite{Unitarity}. 
This bound is seen not as a limit, but  as an
indication that near the perturbative unitarity bound  strong dynamics takes place.

Strong constraints on SM4 can be also obtained from Higgs  searches at the
Tevatron and the LHC. As the dominant mode for Higgs production at hadron colliders is through loop induced gluon-gluon fusion, the Higgs--gluon--gluon vertex ($hgg$) is significantly enhanced by new 
heavy coloured fermions of SM4 in the loop, which couple to the Higgs boson proportionally
to their mass.  The enhancement in the production cross section can be approximated by  a factor of $\displaystyle\frac{\sigma(gg \to h)_{\rm SM4}}{\sigma(gg \to h)_{\rm SM}} \approx 9$ \cite{ggH}. Such 
a large enhancement would certainly increase the event rates for a Higgs signal at experiments and therefore,
non observation of any signal helps in putting strong constraints on the Higgs boson mass in SM4.  
The CDF and D0 experiments exclude a Higgs boson in this scenario for masses $124\;{\rm GeV}\! < \! m_h < 286\;$GeV by considering mainly the $gg\! \to\! h \!\to  \! WW \! \to\!  2\ell 2\nu$ channel
\cite{TeVatron:2012cn}. The LHC experiments recently extended this exclusion  limit up
to $m_h\! \approx \!600$ GeV (at 99\%~CL)  by exploiting also the $gg \! \to  \! h \!
\to ZZ \! \to \! 4\ell, 2\ell 2\nu, 2\ell 2j$ search channels \cite{SM4-LHC_July}. The new data worsens the situation for SM4 \cite{Djouadi:2012ae, Vysotsky:2013gfa}. In addition,  the limits on the low energy phenomenology due to fourth generation fermions in SM4 has also been studied extensively
\cite{UTfit,Alok:2010zj}. 

While there have been many extensive studies of the SM4, there are few analyses of BSM scenarios
with four generations (see however \cite{DePree:2009ed,Aparici:2012vx,Geller:2012tg}).  The reason is that
the fourth generation typically imposes severe restrictions on the models. In particular, there are
difficulties in incorporating a chiral fourth family scenario into any
Higgs doublet model, such as the MSSM \cite{Murdock:2008rx}.  It was
initially shown that due to the large masses for the fourth generation
quarks and large Yukawa couplings, there are no values of
$\displaystyle \tan \beta=\frac{v_u}{v_d}>1$ for which the couplings
are perturbative to the Grand Unification Scale. One would need to invoke different couplings, such as one Higgs doublet only coupling to the fourth generation \cite{Geller:2012tg}. (However, this
condition does not apply  to vector-like quarks \cite{Atre:2011ae}.)
However the MSSM with four generations has received some more attention
\cite{Cotta:2011bu},  as it was shown that for $\tan \beta \simeq 1$ the
model exhibits a strong first order phase transition
\cite{Fok:2008yg}.  Four generations can be incorporated naturally into warped spacetime model \cite{Burdman:2009ih}, seen as perhaps a particular example of composite Higgs, models where the Higgs boson emerges as a condensate of the fourth generation fermions \cite{BarShalom:2010bh}.

Given the serious shortcomings of SM4, we chose to explore the possibility of a four generation model in a simple extension of the SM, the Higgs Triplet Model \cite{Konetschny:1977bn}. This framework which we have chosen to call HTM4, has immediately two advantages:
\begin{enumerate}
\item Unlike Higgs doublet models, there are no problems with Yukawa couplings arising from the ratio of the two doublet vacuum expectation values (VEVs), and thus some of the problems with perturbativity are softened;
\item The Higgs Triplet Model is the simplest scenario to allow for neutrino masses through type-II seesaw mechanism \cite{Perez:2008ha}.
\end{enumerate}
Additionally, we will show that, for small mass splittings within the Higgs multiplets and the additional fermion family, the model satisfies precision conditions on the oblique parameters. We explore whether 
one of the neutral CP-even Higgs with mass $\sim 125$ GeV in the HTM4 could be consistent with the Higgs signals at the LHC.

Our paper is organized as follows. In Section \ref{sec:model} we describe briefly the Higgs Triplet Model with nontrivial mixing and including a fourth generation. We include the oblique corrections and the restrictions imposed on the model in Section \ref{sec:oblique}. Our analysis of the parameter space, including collider and precision electroweak constraints, is discussed in Section \ref{sec:analysis} while we present our results in Section \ref{sec:results} . We summarize and conclude in Section \ref{sec:conclusions}. Additional formulas for production and decay of the Higgs boson(s) are provided in the Appendices (\ref{appendix}).

\section{The Higgs Triplet Model with 4 generations}
\label{sec:model}

The scalar sector of the HTM4 is composed of one isospin doublet field $\Phi$ with 
hypercharge $Y_\Phi=1$ and one triplet field $\Delta$. It is customary to choose the triplet to be a complex field with hypercharge $Y_\Delta=2$.\footnote {A real field with hypercharge $Y_\chi=0$ is also possible:  $$\chi =
\left(
\begin{array}{c}
\chi^+\\
\chi^0 \\
\chi^-
\end{array}\right)\rm{ with }~ \chi^0=\frac{1}{\sqrt{2}}(\chi+v_\chi+i\eta).
$$}
The electric charge is defined to be $Q=T_{3L}+\frac{Y}{2}$, with $T_{3L}$ the third component of the $SU(2)_L$ isospin. The scalar fields $\Phi$ and $\Delta$ can be parameterized as a $1 \times 2$ column, and a $2 \times 2$ matrix, respectively:
\begin{eqnarray}
\Phi=\left[
\begin{array}{c}
\varphi^+\\
\frac{1}{\sqrt{2}}(\varphi+v_\Phi+i\chi)
\end{array}\right],\quad \Delta =
\left[
\begin{array}{cc}
\frac{\Delta^+}{\sqrt{2}} & \Delta^{++}\\
\Delta^0 & -\frac{\Delta^+}{\sqrt{2}} 
\end{array}\right]\rm{ with }~ \Delta^0=\frac{1}{\sqrt{2}}(\delta+v_\Delta+i\eta),
\end{eqnarray}
where $v_\Phi$ and $v_\Delta$ 
are the VEVs of the doublet Higgs field and the triplet Higgs field, respectively, which satisfy 
$v^2\equiv v_\Phi^2+2v_\Delta^2\simeq$ (246 GeV)$^2$. 

The  terms in the Lagrangian relevant for Higgs interactions are given by 
\begin{eqnarray}
\mathcal{L}_{\rm{HTM}}=\mathcal{L}_{\rm{kin}}+\mathcal{L}_{Y}-V(\Phi,\Delta), 
\end{eqnarray}
where $\mathcal{L}_{\rm{kin}}$, $\mathcal{L}_{Y}$ and $V(\Phi,\Delta)$ are 
the kinetic term, Yukawa interaction and scalar potential, respectively. 
The kinetic term for the Higgs fields is 
\begin{eqnarray}
\mathcal{L}_{\rm{kin}\, \Phi, \Delta}&=&(D_\mu \Phi)^\dagger (D^\mu \Phi)+\rm{Tr}[(D_\mu \Delta)^\dagger (D^\mu \Delta)], 
\end{eqnarray}
where the covariant derivatives are defined as
\begin{eqnarray}
D_\mu \Phi=\left(\partial_\mu+i\frac{g}{2}\tau^aW_\mu^a+i\frac{g'}{2}B_\mu\right)\Phi, \quad
D_\mu \Delta=\partial_\mu \Delta+i\frac{g}{2}[\tau^aW_\mu^a,\Delta]+ig'B_\mu\Delta. 
\end{eqnarray}
The fermion composition of the model is augmented by an extra generation of quarks and leptons ($SU(2)_L$ doublets and right-handed singlets):
\begin{eqnarray}
Q^4_{L} =
\left(
\begin{array}{c}
t'\\
b' 
\end{array}\right)_L ,\quad t'_R,~~~b'_R; \quad L^4_{L} =
\left(
\begin{array}{c}
\nu_{\tau'}\\
\tau'
\end{array}\right)_L, \quad \tau_R', ~~~\nu_R'.
\end{eqnarray}
The Yukawa interaction for the Higgs fields is given by 
\begin{eqnarray}
\mathcal{L}_Y&=&-\left[\bar{Q}_L^iY_d^{ij}\Phi d_R^j+\bar{Q}_L^iY_u^{ij}\tilde{\Phi}u_R^j+\bar{L}_L^iY_e^{ij}\Phi e_R^j+\bar{L}_L^\prime Y_{\nu\prime}\tilde{\Phi} \nu^\prime_R+\rm{h.c.}\right] + h_{ij}\overline{L_L^{ic}}i\tau_2\Delta L_L^j+\rm{h.c.},~~~~ \label{nu_yukawa}
\end{eqnarray}
where $\Phi,~\tilde{\Phi}=i\tau_2 \Phi^*$,  $Y_{u,d,e, \nu^\prime}$ are  4$\times$4 complex matrices, and $h_{ij}$ is a $4\times 4$ complex symmetric Yukawa matrix. 
The triplet field $\Delta$ carries lepton number 2.  
The most general form of the Higgs potential involving the doublet $\Phi$ and triplet $\Delta$ under the gauge symmetry is given by \cite{Arhrib:2011uy}
\begin{eqnarray}
V(\Phi,\Delta)&=&m_\Phi^2\Phi^\dagger\Phi+M^2\rm{Tr}(\Delta^\dagger\Delta)+\left[\mu \Phi^Ti\tau_2\Delta^\dagger \Phi+\rm{h.c.}\right]+\frac{\lambda}{4}(\Phi^\dagger\Phi)^2 \nonumber\\
&+&\lambda_1(\Phi^\dagger\Phi)\rm{Tr}(\Delta^\dagger\Delta)+\lambda_2\left[\rm{Tr}(\Delta^\dagger\Delta)\right]^2 +\lambda_3\rm{Tr}[(\Delta^\dagger\Delta)^2]
+\lambda_4\Phi^\dagger\Delta\Delta^\dagger\Phi,~~~~ \label{pot_htm}
\label{eqn:pot}
\end{eqnarray}
where $m_\Phi$ and $M$ are the mass-dimension real parameters, $\mu$ is the lepton-number violating  parameter   of mass-dimension (which can be complex but is taken to be real here), and $\lambda$, 
$\lambda_1$-$\lambda_4$ are dimensionless real coupling constants. 

From the stationary conditions at the vacuum for $(v_\Phi, v_\Delta)$, we obtain $$\frac {\partial V(\Phi,\Delta)}{\partial v_\Phi}=0, \qquad \frac {\partial V(\Phi,\Delta)}{\partial v_\Delta}=0$$  yielding conditions for $m_\Phi^2, M^2$:
\begin{eqnarray}
m_\Phi^2&=&\displaystyle \frac{1}{2}\left[-\frac{v_\Phi^2\lambda}{2}-v_\Delta^2(\lambda_1+\lambda_4)+2\sqrt{2}\mu v_\Delta\right],\\
M^2&=&\displaystyle M_\Delta^2-\frac{1}{2}\left[2v_\Delta^2(\lambda_2+\lambda_3)+v_\Phi^2(\lambda_1+\lambda_4)\right], \label{vc}
\rm{ with }~ M_\Delta^2\equiv \frac{v_\Phi^2\mu}{\sqrt{2}v_\Delta}, 
\end{eqnarray}
which can be used to eliminate $m_\Phi^2$ and $M^2$. 
The mass matrices for the scalar bosons can be diagonalized by rotating the 
scalar fields as 
\begin{eqnarray}
\left(
\begin{array}{c}
\varphi^\pm\\
\Delta^\pm
\end{array}\right)&=&
\left(
\begin{array}{cc}
\cos \beta_\pm & -\sin\beta_\pm \\
\sin\beta_\pm   & \cos\beta_\pm
\end{array}
\right)
\left(
\begin{array}{c}
w^\pm\\
H^\pm
\end{array}\right),\quad 
\left(
\begin{array}{c}
\chi\\
\eta
\end{array}\right)=
\left(
\begin{array}{cc}
\cos \beta_0 & -\sin\beta_0 \\
\sin\beta_0   & \cos\beta_0
\end{array}
\right)
\left(
\begin{array}{c}
z\\
A
\end{array}\right),\nonumber\\
\left(
\begin{array}{c}
\varphi\\
\delta
\end{array}\right)&=&
\left(
\begin{array}{cc}
\cos \alpha & -\sin\alpha \\
\sin\alpha   & \cos\alpha
\end{array}
\right)
\left(
\begin{array}{c}
h\\
H
\end{array}\right),
\end{eqnarray}
where we defined  the mixing angles
\begin{eqnarray}
\tan\beta_\pm=\frac{\sqrt{2}v_\Delta}{v_\Phi},\quad \tan\beta_0 = \frac{2v_\Delta}{v_\Phi}, \quad
\tan2\alpha &=&\frac{4v_\Delta}{v_\Phi}\frac{v_\Phi^2(\lambda_1+\lambda_4)-2M_\Delta^2}{v_\Phi^2\lambda-2M_\Delta^2-4v_\Delta^2(\lambda_2+\lambda_3)}.~~~~~~~~ \label{tan2a}
\label{tan-beta-alpha}
\end{eqnarray}
In addition to the three Goldstone bosons $w^\pm$ and $z$ which give mass to the gauge bosons, 
there are seven physical mass eigenstates $H^{\pm\pm}$, $H^\pm$, $A$, $H$ and $h$. 
The masses of these physical states are expressed in terms of the parameters in the Lagrangian as 
\begin{eqnarray}
m_{H^{++}}^2&=&M_\Delta^2-v_\Delta^2\lambda_3-\frac{\lambda_4}{2}v_\Phi^2,\label{mhpp}\\
m_{H^+}^2&= &\left(M_\Delta^2-\frac{\lambda_4}{4}v_\Phi^2\right)\left(1+\frac{2v_\Delta^2}{v_\Phi^2}\right),\label{mhp}\\
m_A^2 &= &M_\Delta^2\left(1+\frac{4v_\Delta^2}{v_\Phi^2}\right), \label{mA}\\
m_H^2&=&\mathcal{M}_{11}^2\sin^2\alpha+\mathcal{M}_{22}^2\cos^2\alpha-\mathcal{M}_{12}^2\sin2\alpha,\label{mH}\\
m_h^2&=&\mathcal{M}_{11}^2\cos^2\alpha+\mathcal{M}_{22}^2\sin^2\alpha+\mathcal{M}_{12}^2\sin2\alpha,
\end{eqnarray}
where $\mathcal{M}_{11}^2$, $\mathcal{M}_{22}^2$ and $\mathcal{M}_{12}^2$ are 
the elements of the mass matrix $\mathcal{M}_{ij}^2$ for the CP-even scalar states in the $(\varphi,\delta)$ basis which are 
given by
\begin{eqnarray}
\mathcal{M}_{11}^2&=&\frac{v_\Phi^2\lambda}{2},\\
\mathcal{M}_{22}^2&=&M_\Delta^2+2v_\Delta^2(\lambda_2+\lambda_3),\\
\mathcal{M}_{12}^2&=&-\frac{2v_\Delta}{v_\Phi}M_\Delta^2+v_\Phi v_\Delta(\lambda_1+\lambda_4).
\end{eqnarray}

The masses of the $W$ and  $Z$  bosons are obtained at the tree level as 
\begin{eqnarray}
m_W^2 = \frac{g^2}{4}(v_\Phi^2+2v_\Delta^2),\quad m_Z^2 =\frac{g^2}{4\cos^2\theta_W}(v_\Phi^2+4v_\Delta^2).
\end{eqnarray}
The electroweak $\rho$ parameter is defined at tree level as
\begin{eqnarray}
\rho \equiv \frac{m_W^2}{m_Z^2\cos^2\theta_W}= \frac{1+\displaystyle \frac{2v_\Delta^2}{v_\Phi^2}}{1+\displaystyle \frac{4v_\Delta^2}{v_\Phi^2}}. \label{rho_triplet}
\label{eqn:rho}
\end{eqnarray}
As the experimental value of the $\rho$ parameter is near unity, 
$v_\Delta^2/v_\Phi^2$ is required to be much smaller than unity at the tree level. Note in fact than in the HTM4 the $\rho$ parameter is less than $1$, which may contradict the PDG fit \cite{particledata} $\rho_0 = 1.0008^{+0.0017}_{ -0.0007}$ 
obtained from a global fit including the direct
search limits on the standard Higgs boson. However at the 2$\sigma$ level, 
$\rho_0 = 1.0004^{+0.0029}_{-0.0011}$, \cite{particledata}, which is 
compatible with $\delta \rho <0$.  
Relaxing the direct limit on the Higgs mass yields 
$\rho_0 = 1.0008^{+0.0017}_{-0.0010}$, 
again compatible with $\delta \rho <0$, which implies 
an upper bound on $v_\Delta$ of order $2.5-4.6$ GeV. Thus a $v_\Delta \sim {\cal O}(1)$ GeV would safely fit the constraints.
Barring accidental cancellations, the mixing angles are small, the state $h$ behaves mostly as the SM Higgs boson, while 
the other states are almost entirely components of the triplet field. 
Note that the smallness of $v_\Delta/v_\Phi$ insures that the mixing angles $\beta_\pm$ and $\beta_0$ are close to $0$, but given the complex expression defining it, it is worthwhile to note that  $\alpha$, the mixing angle between the doublet and triplet neutral Higgs bosons  remains undetermined \cite{Arbabifar:2012bd}.

In the HTM4, tiny Majorana neutrino masses are generated by the Yukawa interaction with the VEV of the triplet field, which  
is proportional to the lepton number violating coupling constant $\mu$ as
\begin{equation}
(m_\nu)_{ij}=\sqrt{2}h_{ij} v_\Delta=h_{ij}\frac{\mu v_\Phi^2}{M_\Delta^2}. \label{mn}
\end{equation}
If $\mu\ll M_\Delta$ the smallness of the neutrino masses are explained by the type II seesaw mechanism. 

\section{Oblique Corrections}
\label{sec:oblique}
The contributions to the oblique parameters  can in general be written as 
\begin{eqnarray}
S&=&  \frac{4c_W^2s_W^2}{\alpha_{em}} \Bigg[ \frac{\Pi_{\gamma \gamma}^{1PI}(m_Z^2)-\Pi_{\gamma \gamma}^{1PI}(0)}{m_Z^2}+\frac{c_W^2-s_W^2}{c_Ws_W} \frac{\Pi_{Z \gamma}^{1PI}(m_Z^2)-\Pi_{Z \gamma}^{1PI}(0)}{m_Z^2}- \frac{\Pi_{ZZ}^{1PI}(m_Z^2)-\Pi_{ZZ}^{1PI}(0)}{m_Z^2} \Bigg] ,\nonumber  \\
T &=& \frac{1}{\alpha_{em}} \Bigg[\frac{\Pi_{Z Z}^{1PI}(0)}{m_Z^2}-\frac{\Pi_{WW}^{1PI}(0)}{m_W^2}+\frac{2s_W}{c_W}\frac{\Pi_{Z \gamma}^{1PI}(0)}{m_Z^2}
+\frac{s_W^2}{c_W^2}\frac{\Pi_{\gamma \gamma}^{1PI}(0)}{m_Z^2}\Bigg], \nonumber \\
 U &=& \frac{4s_W^2}{\alpha_{em}} \Bigg[s_W^2 \frac{\Pi_{\gamma \gamma}^{1PI}(m_Z^2)-\Pi_{\gamma \gamma}^{1PI}(0)}{m_Z^2}+2c_Ws_W \frac{\Pi_{Z \gamma}^{1PI}(m_Z^2)-\Pi_{Z \gamma}^{1PI}(0)}{m_Z^2}+c_W^2 \frac{\Pi_{ZZ}^{1PI}(m_Z^2)-\Pi_{ZZ}^{1PI}(0)}{m_Z^2}  \nonumber \\ 
  && \hspace*{0.5in} - \frac{\Pi_{WW}^{1PI}(m_W^2)-\Pi_{WW}^{1PI}(0)}{m_W^2}\Bigg]
 \end{eqnarray}

The expressions for  $ \Pi_{ZZ}^{1PI}(p^2), \Pi_{Z\gamma}^{1PI}(p^2), \Pi_{WW}^{1PI}(p^2) $ and $\Pi_{\gamma \gamma}^{1PI}(p^2)$ for the HTM are given in terms of Passarino-Veltman functions in Appendix B of ref. \cite{Aoki:2012jj}, and the the contributions to the oblique corrections due to the  sequential fourth generation can be found in ref. \cite{He:2001tp}. Note that both the new scalar sector 
(with nontrivial mixing amongst themselves and its modified couplings to the weak gauge bosons)  and the fourth generation fermions
will lead to significant and nonvanishing contributions to the oblique electroweak corrections. It is well 
known that in the limit of degenerate isospin multiplets (in this case the fourth generation family of leptons 
and quarks) there is a positive contribution $\Delta S=0.21$ which can now significantly be altered once the 
HTM contributions are included. In fact, as we show later through our analysis, the HTM contributions 
cancel the large positive contributions coming from the fourth generation even with the degeneracy not lifted between the isodoublets. 
We also find that a large mass splitting in the fermion isodoublets which would otherwise have not been preferred because of the large $\Delta T$ corrections can still be allowed once the HTM and SM4 contributions are combined.    
%%%%%%

\section{Analysis Framework}
\label{sec:analysis}
%%%%
The main motivation in this analysis is to salvage the fourth generation chiral fermions. The HTM4 model contains, in addition to the SM 
particle content, a complex triplet scalar along with  a sequential fourth generation of chiral fermions. The free parameters in the theory are the fourth generation fermion masses (we neglect the CKM mixings), 
the independent coefficients in the potential of the Higgs sector (see Eq. \ref{eqn:pot}) and the VEV's for the neutral components of the scalar multiplets. 
To limit the available number of parameters we have used the various constraints on the coefficients of the Higgs potential (Section \ref{appendix2}). This reduces the 
effective number of free parameters. In addition, we consider both the masses of $\tau^{\prime}$ and $\nu_{\tau^{\prime}}$ to be fixed at 
$250$ GeV. The reason is the following: the strongest constraint on the fourth generation lepton masses comes from the 
$Z$-boson width, forcing the fourth neutrino to be heavier than $M_Z/2$. The most stringent bounds on the fourth generation lepton masses come from L3 at LEP \cite{L3},  obtained from $e^+e^-$ collisions at $\sqrt{s}=200$ GeV:
\begin{equation}
m_{l^\prime}>110.8~{\rm GeV}, \qquad m_{\nu^\prime} >\left \{\begin{array}{ll} 90.3~{\rm GeV~for~ Dirac}~\nu\\80.5~{\rm GeV~for~ Majorana}~\nu, \end{array} \right.
\end{equation}
rendering our choice of parameters quite conservative. We also note that large electroweak corrections for 
the $h\to gg$ decay mode can be moderated to small values with the above choice (or lower) of the fourth generation lepton masses \cite{Passarino:2011kv}.

The method of our analysis is roughly based on the following steps. Using HIGLU \cite{Spira:1995mt}, we calculate the NNLO cross 
sections for Higgs production in the gluon fusion channel, obtained by varying the fourth generation quark masses between $550$ GeV and $750$ 
GeV. Our choice of masses is based on the results of searches by CMS and ATLAS. But we allow for a softening of the limits as these are obtained 
using SM4 specific assumptions. Direct searches at ATLAS and CMS always assume that a single specific decay has 100\% branching ratio, usually 
$t^\prime \to b W$ and $b^\prime \to t W$. While constraints on the mixing between the first and second generations with 
the fourth generation can be extracted  from flavor-changing neutral current bounds in $K^0-{\bar K}^0$ and $D^0-{\bar D}^0$, the size of the CKM4 matrix 
elements is still allowed to be larger than the smallest matrix elements in the usual CKM matrix \cite{Kribs:2007nz}, 
$|V_{ub'}|,~|V_{t'd}|,~|V_{cb'}| \simlt 0.04$. Additionally, the mixing between the third and fourth generations is very weakly 
constrained, as  there is a weak limit from the single top production, $|V_{tb}|>0.89\pm 0.07$ \cite{particledata}, and this bound is 
stronger than that obtained using the electroweak precision data. Though, while it seems likely that the fourth generation quarks will decay into the third generation, this need not be so, and not with 100\% branchings, which would modify the limits on the masses.  The fourth generation fermions can also decay to a Higgs boson via flavour violating couplings to Higgs, which can be introduced by including certain higher dimensional operators \cite{Harnik:2012pb}. This can further lead to weakening of the constraints on the fourth generation quark masses. These considerations allow us to vary the mass of 
fourth generation quarks from values much below the current experimental bounds,  which assume very specific decay patterns for the heavy quarks in their analyses.

We find the (gluon fusion) production cross section to be $\sim195$ pb ($\sim153$ pb) at 8 TeV (7 TeV) which is about a factor of 10 larger than its 
SM expectation. This enhancement is known, and is one of the reasons why the fourth generation chiral fermions is ruled out without  additional 
new physics effects. Next, we calculate the decay widths of the Higgs boson in its various decay modes using {\tt HDECAY} \cite{Djouadi:1997yw}. Here also we vary  the masses of the fourth generation  quark masses in the 550-750 GeV range. We include the electroweak corrections  (EW) coming from the fourth generation contributions in all the tree level decay processes as well as the loop induced gluon mode. However we 
have chosen to ignore the electroweak corrections to  $h \to \gamma\gamma,~Z\gamma$  decay widths in our analysis, as there are additional particles ($H^\pm ~{\rm and}~ H^{\pm\pm}$) contributing in the loop 
and therefore the EW contributions will be severely altered from what is computed in the literature. While 
the electroweak radiative corrections to the $gg\to h$ process are significant \cite{Djouadi:2012ae}, for a specific choice of fermion masses,  $m_{b'}\! = \! m_{t'}+50\;{\rm GeV}= \! m_{\ell'} \! = \! m_{\nu'}
\sim 600 $ GeV,  they lead to an increase (decrease) of the cross section at low
(high) Higgs masses, $M_h\!\approx\! 120\;(600)$ GeV, by $\approx 12\%$ \cite{Kribs:2007nz}.  We also note that the EW corrections in the $gg \to h$ mode can be kept within 5\% with appropriate choice for the mass of the fourth generation leptons \cite{Passarino:2011kv}. However, for the decays $h \to f {\bar f}$ and $h \to V V$, the ${\cal O}(G_F m^2_{f'})$  terms,  implemented by multiplying
the couplings $g_{hXX}$ by the electroweak correction term $1+\delta^X_{ew}$ has been included in 
{\tt HDECAY} \cite{Djouadi:1997yw, Bredenstein:2006ha}. Results from the precise calculations indicate that the approximation of including only the leading terms also works very well in this case \cite{Djouadi:2012ae}.

We then separately scan the parameter space including the Higgs and the electroweak sector. We impose the existence of a  boson of mass $\sim$ 125 
GeV, consistent with the particle discovered at the LHC, and which has properties similar to the SM Higgs (whether it is the SM Higgs or not will only be revealed by 
knowing its properties to a more accurate extent). In our model, which has two CP-even neutral scalar Higgs bosons, we constrain the 
mass of one of these to be in the range 123-127 GeV. We keep this small window in the mass in order to account for the uncertainties in the 
mass measurements in the various channels of the Higgs decay. We constrain the second CP-even neutral Higgs to be in the range 97-99 
GeV. Our inspiration in choosing a Higgs boson of such a mass is the 2.3$\sigma$ excess seen at 98 GeV by LEP  \cite{Schael:2006cr,Barate:2003sz}, which has still not been 
ruled out experimentally and is outside LHC and Tevatron sensitivity. From the parameters $v_{\Delta}$, $\mu$, $\lambda$, $\lambda_{1}$, $\lambda_{2}$, $\lambda_{3}$ and 
$\lambda_{4}$, we can calculate the masses of the two neutral CP-even Higgs bosons, $m_{h}$ and $m_{H}$, the CP-odd Higgs, $m_{A}$ and of 
the four charged Higgs bosons, $m_{H^{\pm}}$ and $m_{H^{\pm \pm}}$. From the above parameters we also obtain the four mixing angles, 
$\beta_{0}$, $\beta_{\pm}$ and $\alpha$. (See Section \ref{appendix1} where we give the formulas relating the above seven parameters to the Higgs masses and the four mixing angles). 
%In our study, we fix the value of $v_{\Delta}$ to a value consistent with  the $\rho$ parameter (see Eq. \ref{eqn:rho}) lying within its 3$\sigma$ uncertainty. Also, 
To reduce the enhanced cross section of the Higgs ($h$) 
production in the gluon fusion channel, we let $\cos^{2}\alpha$ vary between $1/12$ and $1/2$\footnote{Naively one might expect that a 
$\cos^{2}\alpha$ $\sim1/10$ will compensate for the $\mathcal{O}(10)$ enhancement in the gluon-fusion initiated cross section. But 
as is evident from some of the decay widths, the $\alpha$ enters in a nontrivial manner and not just as an overall 
$\cos^{2}\alpha$ factor. Hence, we keep this range instead of a single fixed value because our aim is to ensure that the value of $\sigma \times$ Branching ratio (BR) 
is close to its SM counterpart.}. In addition, we let both $\mu$ and $\lambda$ to vary, such that  the constraints on their magnitudes (as given in section~\ref{appendix2}) are satisfied. We allow $\lambda_{1}$ to vary around a small positive number, but do not 
impose an upper bound. We apply a simplifying assumption by choosing $\lambda_{2}$ = $\lambda_{3}$. Also, we choose negative values of $\lambda_{4}$, yielding the mass hierarchy $m_{H^{\pm \pm}} > m_{H^{\pm}}$.  
(Should we have chosen $\lambda_{4}$ to be positive, we would have obtained a different hierarchy in the masses of the charged Higgs bosons). 

To check with the current Higgs data and see what portion of the parameter space is allowed, we construct the theoretical signal 
strength in the $i^{th}$ channel as defined below:
\begin{equation}
 \mu^{i} = \frac{R_{i}^{prod} \times R_{i}^{decay}}{R^{width}}
\end{equation}
where $\mu^{i}$ is the theoretically computed signal strength and $R_{i}^{prod}$, $R_{i}^{decay}$ and $R^{width}$ are the factors 
modifying the production cross section, decay width in the $i^{th}$ channel and the total decay width respectively. The signal 
strengths provided by the experimental collaborations are given by $\hat{\mu}^{i} = \sigma_{i}^{obs}/\sigma_{i}^{SM}$, with 
their 1$\sigma$ uncertainties given by $\sigma_{i}$. Here $\sigma_{i}^{obs}$ denotes the observed signal cross section for a 
particular Higgs mass, whereas $\sigma_{i}^{SM}$ is the signal cross section for an SM Higgs boson of the same mass. In order to obtain
$\mu^{i}$ for each decay channel, we compute the modifications in the production cross sections from the various modes, the 
modifications in the partial decay widths in the various channels, as well as the change in the total decay width. The modifications in 
the production cross sections are as follows:
\begin{itemize}
 \item The modification in the gluon fusion production channel, $R_{GF}$ is a function of the masses of the fourth generation chiral fermions and 
       the mixing angle $\alpha$.
 \item The factors modifying the production cross sections in the weak boson fusion channel or in the associated production mode, viz.
       $R_{VBF}$, $R_{WH}$ and $R_{ZH}$ are functions of $v_{\phi}$, $v_{\Delta}$ and $\alpha$.
 \item The modification in the $t\bar{t}h$ production mode, $R_{t\bar{t}h}=\cos^{2}\alpha$.
\end{itemize}

Similarly, the factors modifying the decay widths are:
\begin{itemize}
 \item $R_{h \rightarrow \gamma \gamma} = \frac{\bigg| \sum \limits_{f=t,b,\tau,t',b',\tau'} N^f_c Q_f^2 g_{h ff} A_{1/2}^{h} (\tau_f) + g_{h WW} A_1^{h} (\tau_W) + {g}_{h H^\pm\,H^\mp} A_0^{h}(\tau_{H^{\pm}})+ 4 {g}_{h H^{\pm\pm}H^{\mp\mp}}A_0^{h}(\tau_{H^{\pm\pm}}) \bigg|^2}{\bigg| \sum \limits_{f=t,b,\tau} N^f_c Q_f^2 g_{h ff} A_{1/2}^{h} (\tau_f) + g_{h WW} A_1^{h} (\tau_W)\bigg|^{2}}$ \\ \\
       Please see section \ref{subsec:decay} for more details on the formulas.
 \item $R_{h \rightarrow Z Z^{*}} = ({v_\phi}\cos\alpha+4v_\Delta \sin\alpha)^2/{v^{2}}$.

 \item $R_{h \rightarrow W W^{*}} = ({v_\phi}\cos\alpha+2v_\Delta \sin\alpha)^2/{v^{2}}$.

 \item After computing the fermionic decay widths using {\tt HDECAY} by varying the fourth generation 
quark masses, we further modify them by a factor $(\cos \alpha / \cos \beta_{\pm})^{2}$. We then 
divide these by the corresponding SM decay widths to obtain 
       $R_{h \rightarrow b \bar{b}}$ and $R_{h \rightarrow \tau \bar{\tau}}$.
 \item We also compute $R_{h \rightarrow g g}$, $R_{h \rightarrow \mu^{+} \mu^{-}}$, $R_{h \rightarrow c \bar{c}}$, 
       $R_{h \rightarrow s \bar{s}}$ and $R_{h \rightarrow Z \gamma}$ in order to compute $R^{width}$, i.e. the modification in the total decay width of $h$.
       
\end{itemize}

After constructing the $\mu^{i}$s for the full parameter space, we compare these with the 2$\sigma$ allowed ranges of 
$\hat{\mu}^{i}$ as given by the experimental collaborations (see Table~\ref{tab:tab1})\footnote{We deliberately avoid an involved 
statistical analysis (such as $\chi^2$) because of the large number of parameters and limited number of data points as we have only considered the 
inclusive signal strengths (apart from the $b\bar{b}$ channel where the results correspond to the associated production mode only) for 
the various channels given by ATLAS and CMS \cite{Banerjee:2012xc}.}. This gives us a reduced parameter space. We then proceed to check if this reduced 
parameter space satisfies the $S$, $T$ and $U$ bounds within the 1$\sigma$ uncertainty limits \cite{Baak:2012kk}. For consistency, we have checked our results for oblique parameters against that in 
the SM \cite{Hagiwara:1994pw}.  
We performed our analysis  for two benchmark values of $v_{\Delta}$ ($v_{\Delta}$ $=$ $1$, $3$ GeV). We found that, after imposing all conditions, we could not find any surviving parameter points for $v_{\Delta}=1$ GeV within the 1$\sigma$ range for $\Delta S$ and $\Delta T$ combined, although a significant region is allowed 
by the Higgs data. Thus the model is restricted to larger values of $v_{\Delta}$. However we find that at  3$\sigma$ allowed ranges of $\Delta S$ and $\Delta T$, lower values of $v_{\Delta} \simeq 1$ GeV could still give us an allowed region in the parameter space. Thus we may claim that our choice  for the allowed range in the $S$, $T$ and $U$ planes for oblique corrections is slightly more restrictive.

We must add that in our analysis, we took the same cut-efficiencies for all the production channels while computing the signal strengths. This should not affect our results as the cross section coming from the gluon-fusion production mode dominates significantly over the other production modes. The following formulas were used in combining the theoretically computed $\mu^i$ values for the 
$Wh \rightarrow l\nu b \bar{b}$, $Zh \rightarrow l^{+} l^{-} b \bar{b}$ and 
$Zh \rightarrow \nu \bar{\nu} b \bar{b}$:
\begin{align}
 \frac{1}{\bar{\sigma}^2}=\sum \limits_{i} \frac{1}{\sigma_{i}^2}, \nonumber \\
 \frac{\bar{\mu}}{\bar{\sigma}^2}=\sum \limits_{i} \frac{\mu_{i}}{\sigma_{i}^2}.
\end{align}
These yield the combined 1$\sigma$ uncertainties and the combined signal strengths. Since the 
experimental collaborations have reported a single signal strength for the $h \rightarrow b \bar{b}$ channel in the associated production mode, such combinations have to be included. 
\begin{table}[t]
\centering
\begin{tabular}{|l|c|c|c|}
\hline
Channel & $\hat{\mu}$ & Experiment & Energy in TeV (Luminosity in fb$^{-1}$)\\
\hline
\hline
  $h \rightarrow \gamma \gamma$ & $1.55^{+0.33}_{-0.28}$ & ATLAS & $7~(4.8)$ + $8~(20.7)$ \\
\hline
  $h \rightarrow \gamma \gamma$ & $0.78^{+0.28}_{-0.26}$ & CMS & $7~(5.1)$ + $8~(19.6)$ \\
\hline
  $h \xrightarrow{Z Z^{*}} 4l$ & $1.43_{-0.35}^{+0.40}$ & ATLAS & $7~(4.6)$ + $8~(20.7)$ \\
\hline
  $h \xrightarrow{Z Z^{*}} 4l$ & $0.93_{-0.25}^{+0.29}$ & CMS & $7~(5.1)$ + $8~(19.7)$ \\
\hline
  $h \xrightarrow{W W^{*}} 2l 2\nu$ & $0.99_{-0.28}^{+0.31}$ & ATLAS & $7~(4.6)$ + $8~(20.7)$ \\
\hline
  $h \xrightarrow{W W^{*}} 2l 2\nu$ & $0.72_{-0.18}^{+0.20}$ & CMS & $7~(4.9)$ + $8~(19.4)$ \\
\hline
  $h \rightarrow b \bar{b}$ & $0.20_{-0.60}^{+0.70}$ & ATLAS (VH) & $7~(4.7)$ + $8~(20.3)$ \\
\hline
  $h \rightarrow b \bar{b}$ & $1.00_{-0.50}^{+0.50}$ & CMS (VH) & $7~(5.1)$ + $8~(18.9)$ \\
\hline
  $h \rightarrow \tau \bar{\tau}$ & $1.4_{-0.40}^{+0.50}$ & ATLAS & $8~(20.3)$ \\
\hline
  $h \rightarrow \tau \bar{\tau}$ & $0.78_{-0.27}^{+0.27}$ & CMS & $7~(4.9)$ + $8~(19.7)$ \\
\hline
  $h \xrightarrow{W W^{*}} 2l 2\nu$ & $1.4^{+0.70}_{-0.60}$ & ATLAS (VBF) & $7~(4.6)$ + $8~(20.7)$ \\
\hline
  $h \xrightarrow{W W^{*}} 2l 2\nu$ & $0.62^{+0.58}_{-0.47}$ & CMS (VBF) & $7~(4.9)$ + $8~(19.4)$ \\
\hline
\hline
 
\end{tabular}
\caption{Data set used in our analysis, with the values of $\hat{\mu_i}$
  in various channels and their 1$\sigma$ uncertainties as reported by the
  ATLAS \cite{Aad:2013wqa, ATLAS:tautau} and CMS collaborations \cite{CMS-PAS-HIG-13-001, Chatrchyan:2013iaa, Chatrchyan:2013mxa, Chatrchyan:2014vua}.}
\label{tab:tab1}
\end{table}

%%%%%%%%%%%%%%%%%%%%%%%%%%%%%%%%%%%%%%%%%%%%%%%%%%%%%%%%
\section{Results}
\label{sec:results}
\begin{figure}[t!]
\centering
\includegraphics[width=3.1in,height=2.3in]{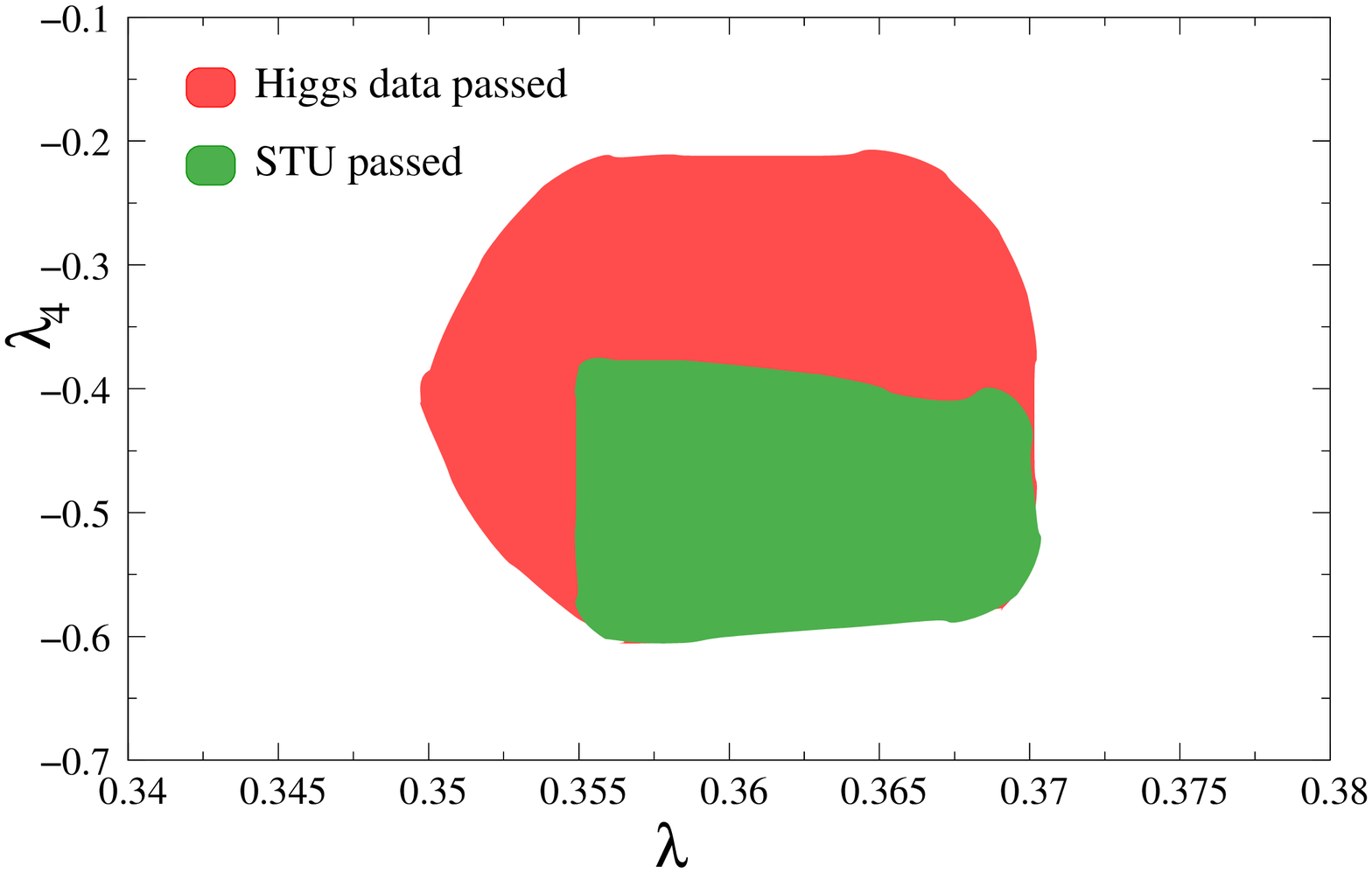}
\includegraphics[width=3.1in,height=2.3in]{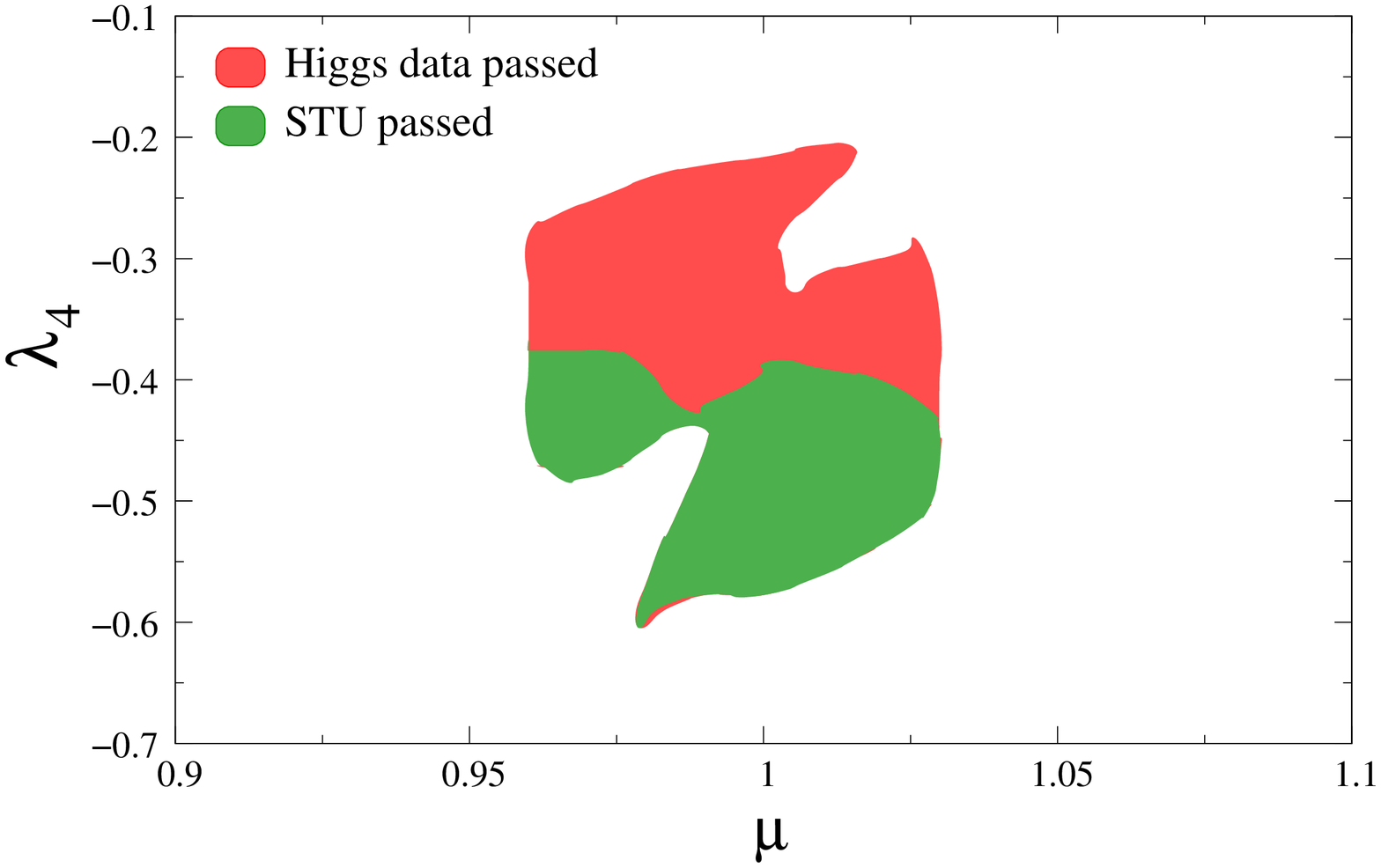}
\includegraphics[width=3.4in,height=2.3in]{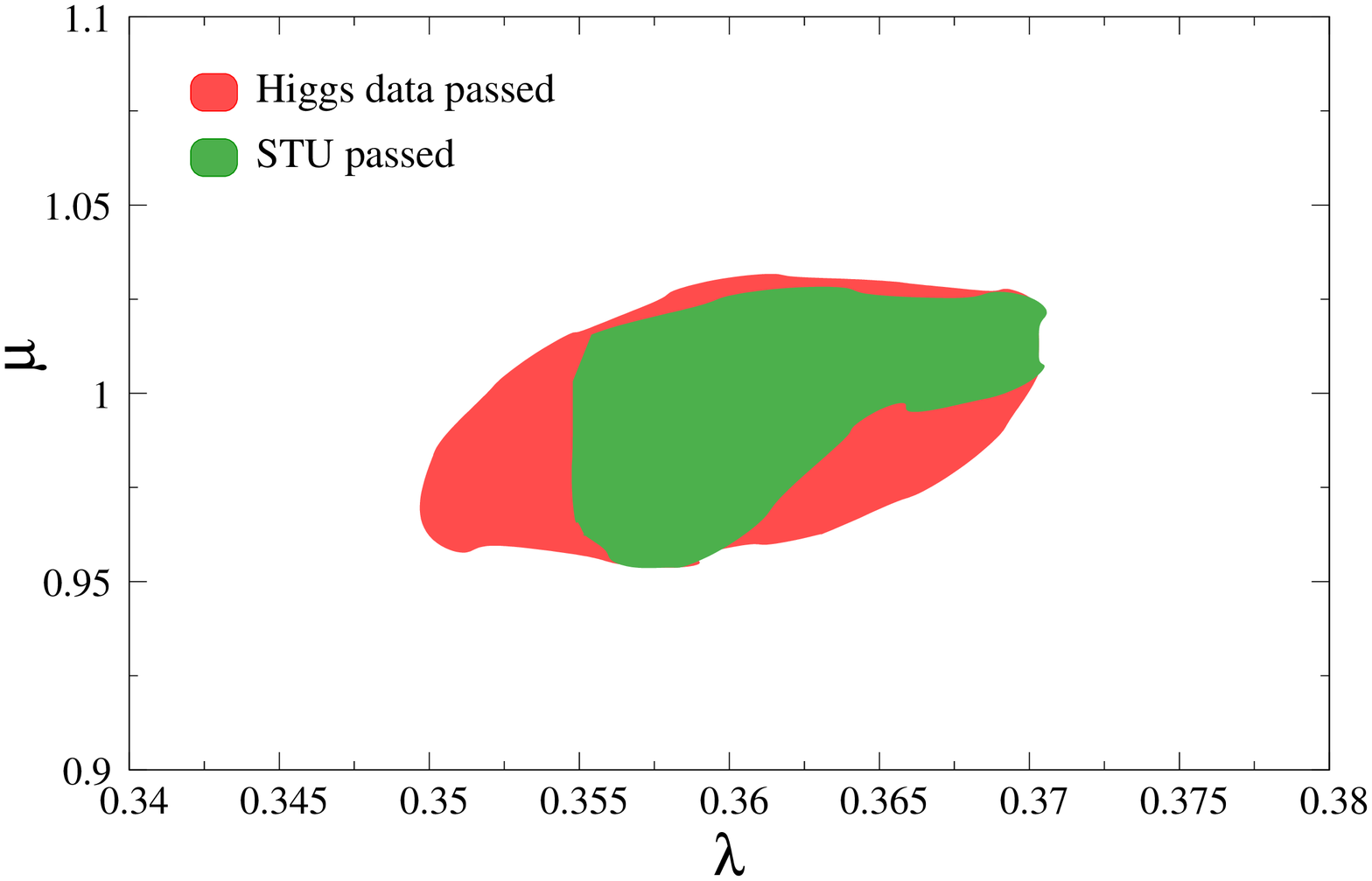}
\caption{\small Scatter plots for the allowed regions in the parameter space : $\lambda_{4}$ as a function of $\lambda$ (top left); 
$\lambda_{4}$ as a function of $\mu$ (top right); $\mu$ as a function of $\lambda$ (bottom). The full shaded (red+green) regions are allowed by the Higgs data while only the green region is allowed by the $S,T$ and $U$ constraints at 1.25$\sigma$. In all the plots, apart from the axes shown, all the other parameters were also varied over their allowed ranges. Note that we varied $m_{t'}, m_{b'} \simeq 550-600$ GeV while higher masses were disallowed
from the Higgs data within 2$\sigma$ standard deviations.}
\label{lam4-mu-lam1}
\end{figure}

We now present our results for the parameter scan of the HTM4 which is consistent with the Higgs data and the oblique correction constraints. We find that there is a significant region in the parameter space which ensures that the otherwise constrained model for fourth generation chiral fermions is still allowed if we consider the effects of a triplet Higgs sector, provided it has a nontrivial mixing with the scalar doublet in 
the SM.  
We have illustrated our results through some scatter plots for the various coefficients of the scalar potential
 giving rise to the aforementioned nontrivial mixing in the Higgs sector and gives us a $\sim 125$ GeV scalar consistent with the LHC data. In Fig.~\ref{lam4-mu-lam1}, we show the allowed parameter 
space in the $\lambda_{4}-\lambda$, $\lambda_{4}-\mu$ and $\mu-\lambda$ planes, for  $m_{t'},m_{b'}$  lying in the region $550-600$ GeV. 
%%%%%%%
%%%%%%
Note that although our scan over the fourth generation quark masses
is between $550-750$ GeV, the Higgs data seems to constrain more severely the fourth generation quark mass above 600 GeV,
at least within the 2$\sigma$ uncertainty limits of all respective $\hat{\mu}^{i}$. This is found to happen because 
the EW corrections to the $h \to ZZ^*, WW^*$ reduce the branching fractions of these modes by a large amount for higher values of the quark masses, which therefore affects the Higgs data significantly. We therefore
allow for a 3$\sigma$ deviation in the experimentally observed signal strengths and find that we do get a viable region in the parameter space for the higher values of  the fourth generation quark masses.  

%%%%%%%
\begin{figure}[t!]
\centering
\includegraphics[width=2.9in,height=2.1in]{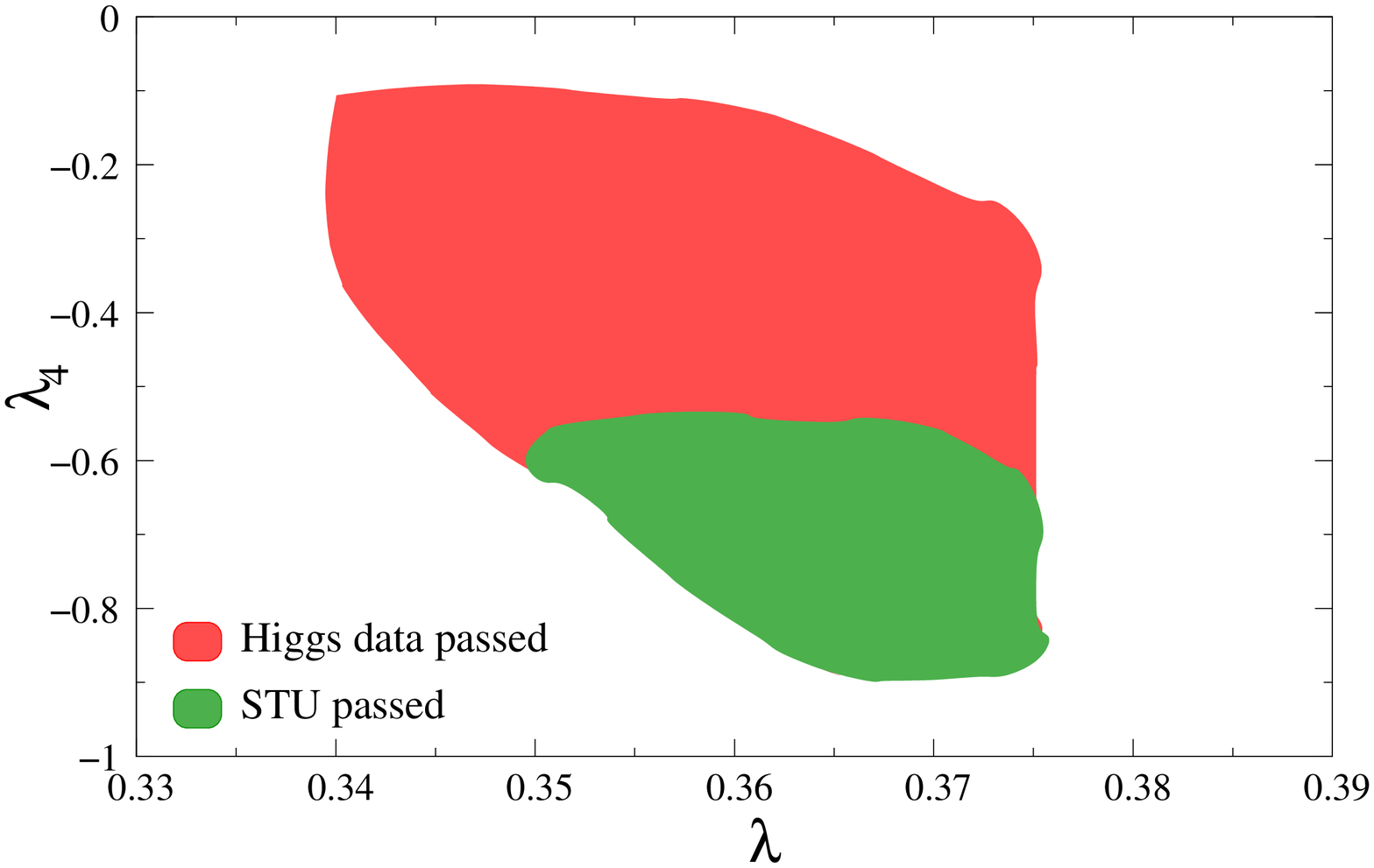}
\includegraphics[width=2.9in,height=2.1in]{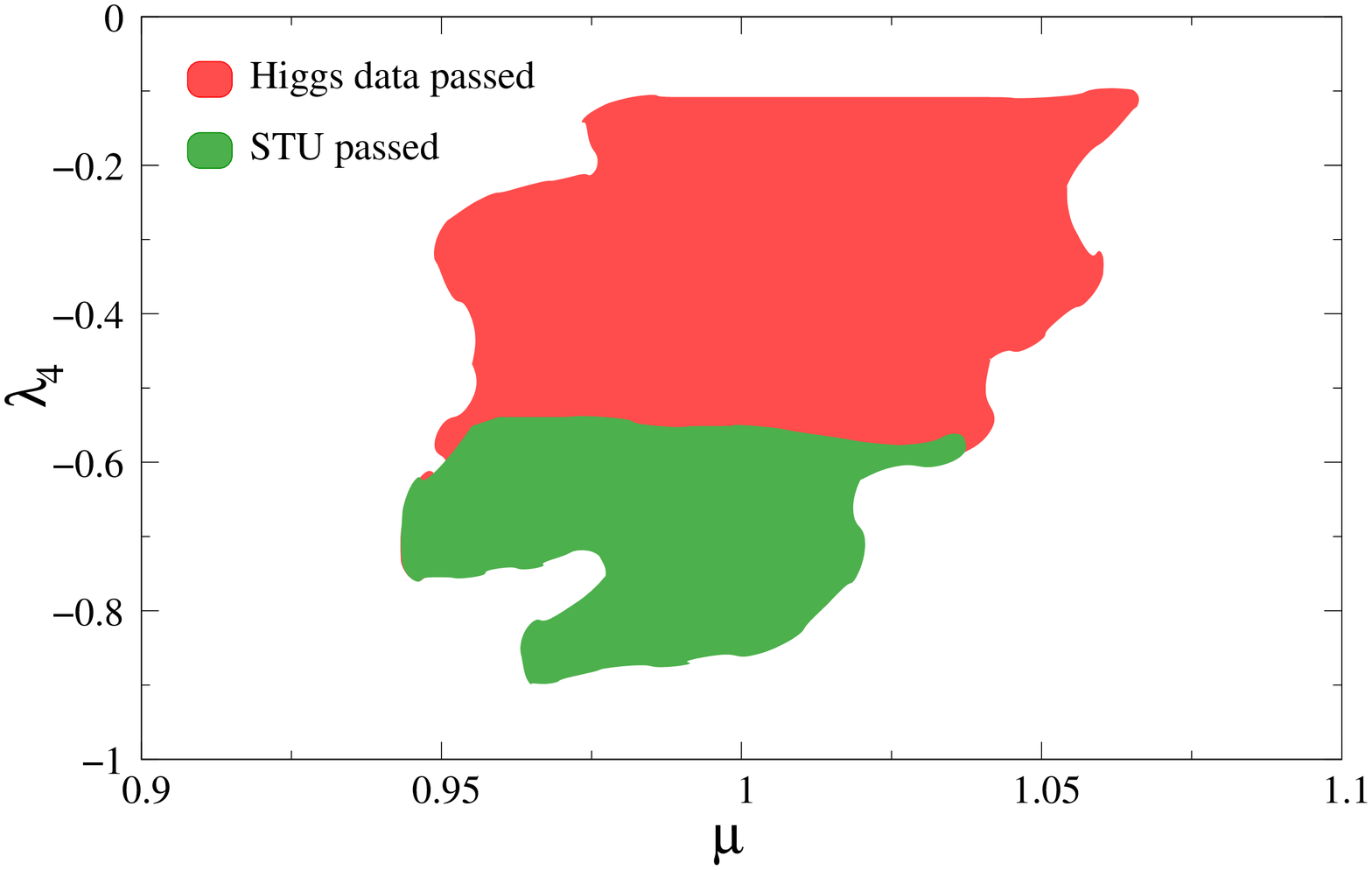}
\includegraphics[width=3.4in,height=2.1in]{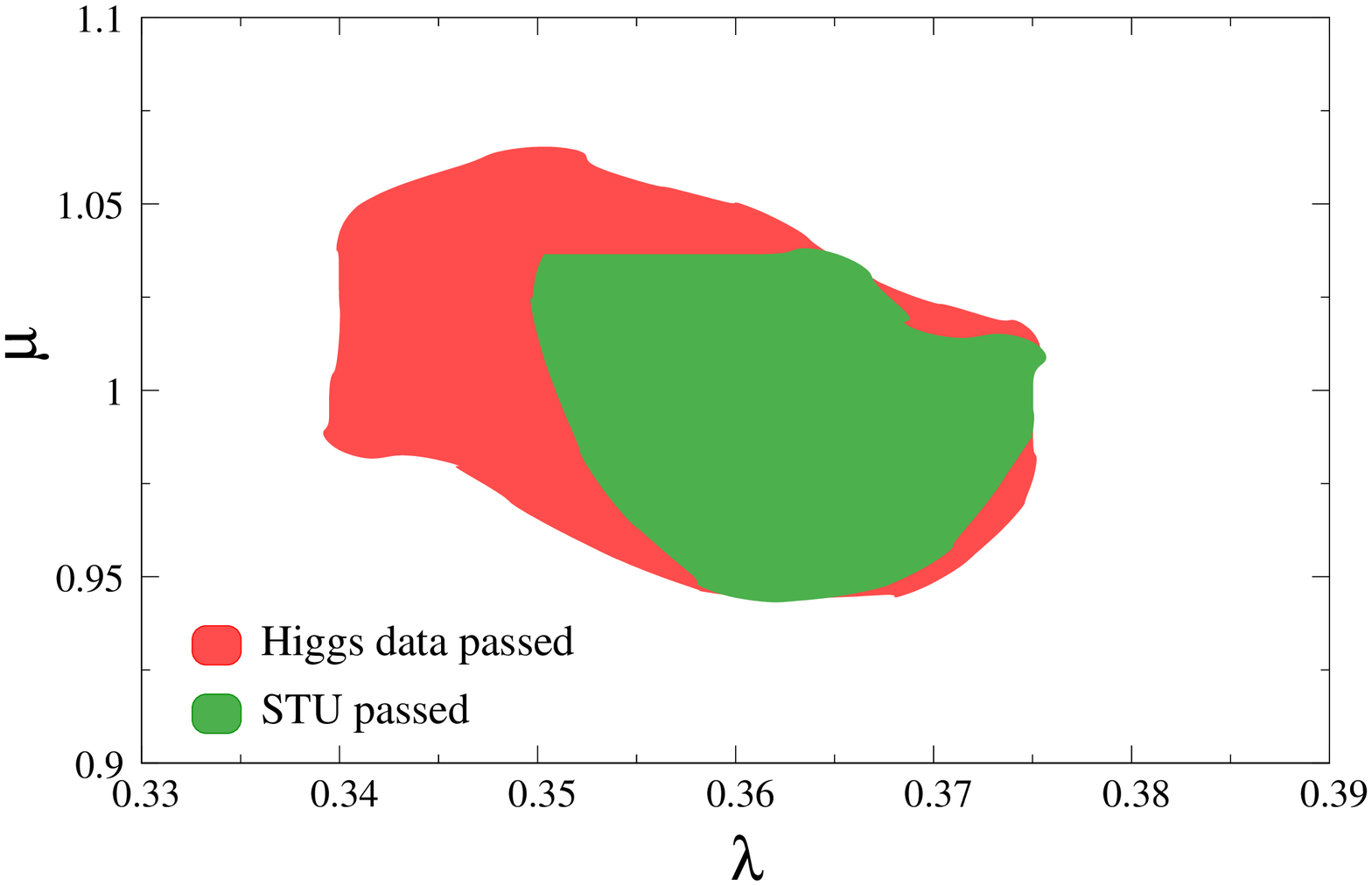}
\caption{\small Scatter plots for the allowed regions in the parameter space : $\lambda_{4}$ as a function of $\lambda$ (top left); 
$\lambda_{4}$ as a function of $\mu$ (top right); $\mu$ as a function of $\lambda$ (bottom). The full shaded (red+green) regions are allowed by the Higgs data while only the green region is allowed by the $S,T$ and $U$ constraints at 1$\sigma$. In all the plots, apart from the axes shown, all the other parameters were also varied over their allowed ranges. In all the plots, apart from the axes shown, all the other parameters were also varied over their allowed ranges. Note that here we have varied $m_{t'}, m_{b'} \simeq 550-750$ GeV while the allowed region from the Higgs data was increased to lie 
within 3$\sigma$ errors.}
\label{lam4-mu-lam2}
\end{figure}
%%%%%% 
We must note that, in all of these plots, the other parameters are also varied. This way, we have shown the projection of the allowed parameter space in the above three planes. We find 
that $\lambda$ has an allowed range of $0.35$-$0.37$ and $\mu$ has an approximate allowed range of $0.96$-$1.03$, when we consider the 2$\sigma$ Higgs data constraints (as explained in the previous section).  
On the other hand, given our choice of negative $\lambda_{4}$, the allowed region from the Higgs data varies between $-0.6$ and $-0.2$. But as we can see from each of these plots, when we calculate the oblique corrections in the HTM4 model and impose the $S$, $T$ and $U$ constraints at 1.25$\sigma$\footnote{We find that for 1$\sigma$ constraints in $S,T$ and $U$, we do not get any viable parameter points which also satisfy the Higgs data within 2$\sigma$ standard deviations.} deviations from the respective central values, the allowed region reduces considerably. It is worth noting that we have been conservative in allowing only 1.25$\sigma$ deviation in $\Delta S$ and $\Delta T$ values,  and that we are still able to achieve a viable parameter space with larger mass splittings in the fourth generation quark masses allowed by oblique corrections when compared to the conventional SM4. In fact, we find that for 
non-degenerate masses 
for $\nu_{\tau'}$ and $\tau'$ (allowing them to have a certain mass splitting) it is possible to obtain a significantly larger parameter space still allowed by the oblique parameter constraints. So, apart from 
varying the 
fourth generation quark masses, we could in principle have varied the fourth generation lepton masses also (while keeping the EW corrections small) and obtained a larger parameter space.

If we increase the masses of the fourth generation quarks by allowing them to now vary in the region $550-750$ GeV, the region 
of the parameter space that survives oblique corrections ($1\sigma$) and the Higgs data ($3\sigma$) changes.  In Fig.~\ref{lam4-mu-lam2}, we revisit  the allowed parameter 
space in the $\lambda_{4}-\lambda$, $\lambda_{4}-\mu$ and $\mu-\lambda$ planes for heavier fourth generation masses. As before, in all of these plots the other 
%%%
\begin{figure}[t!]
\centering
\includegraphics[width=3.1in,height=2.8in]{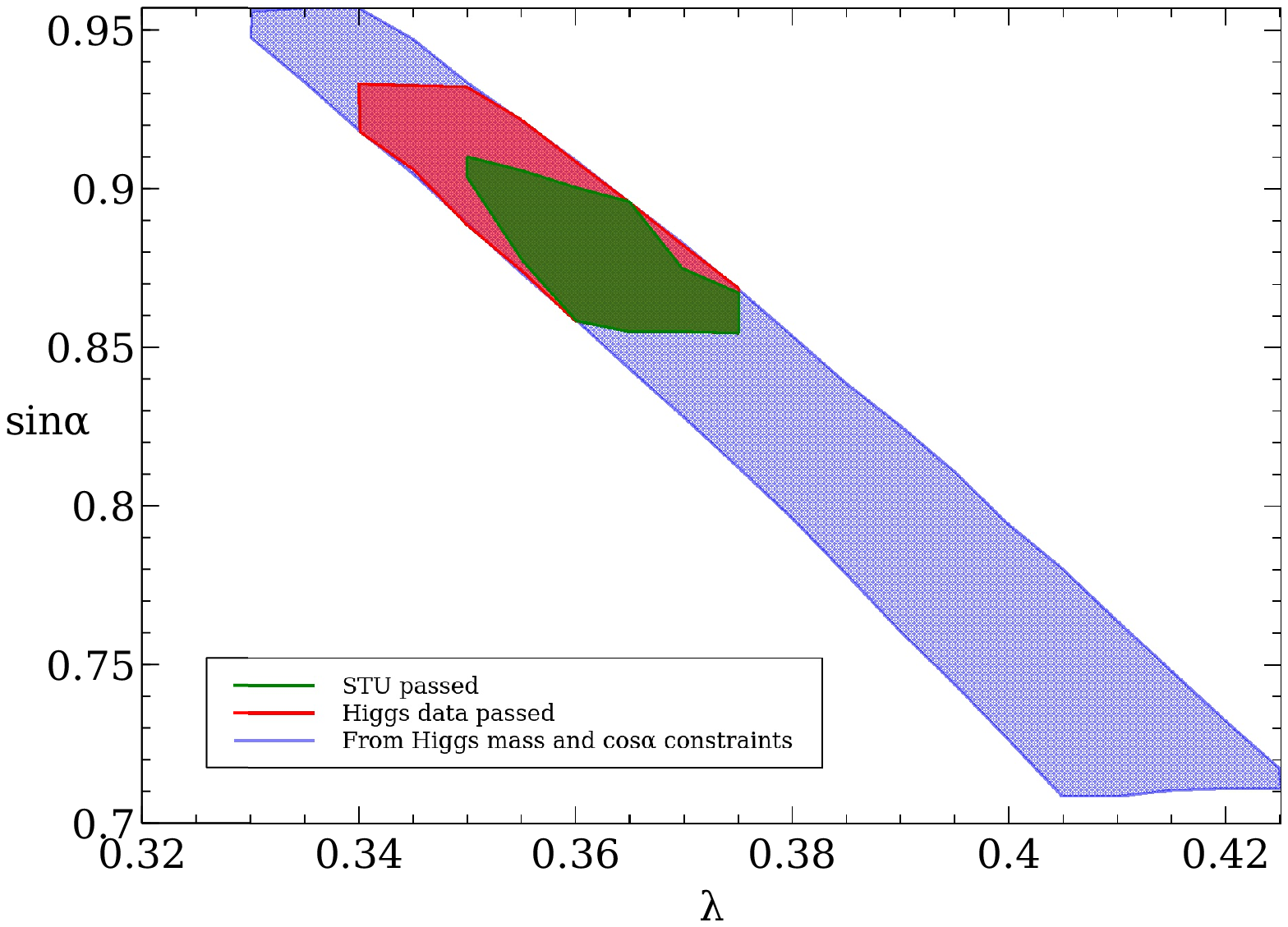}
\includegraphics[width=3.1in,height=2.8in]{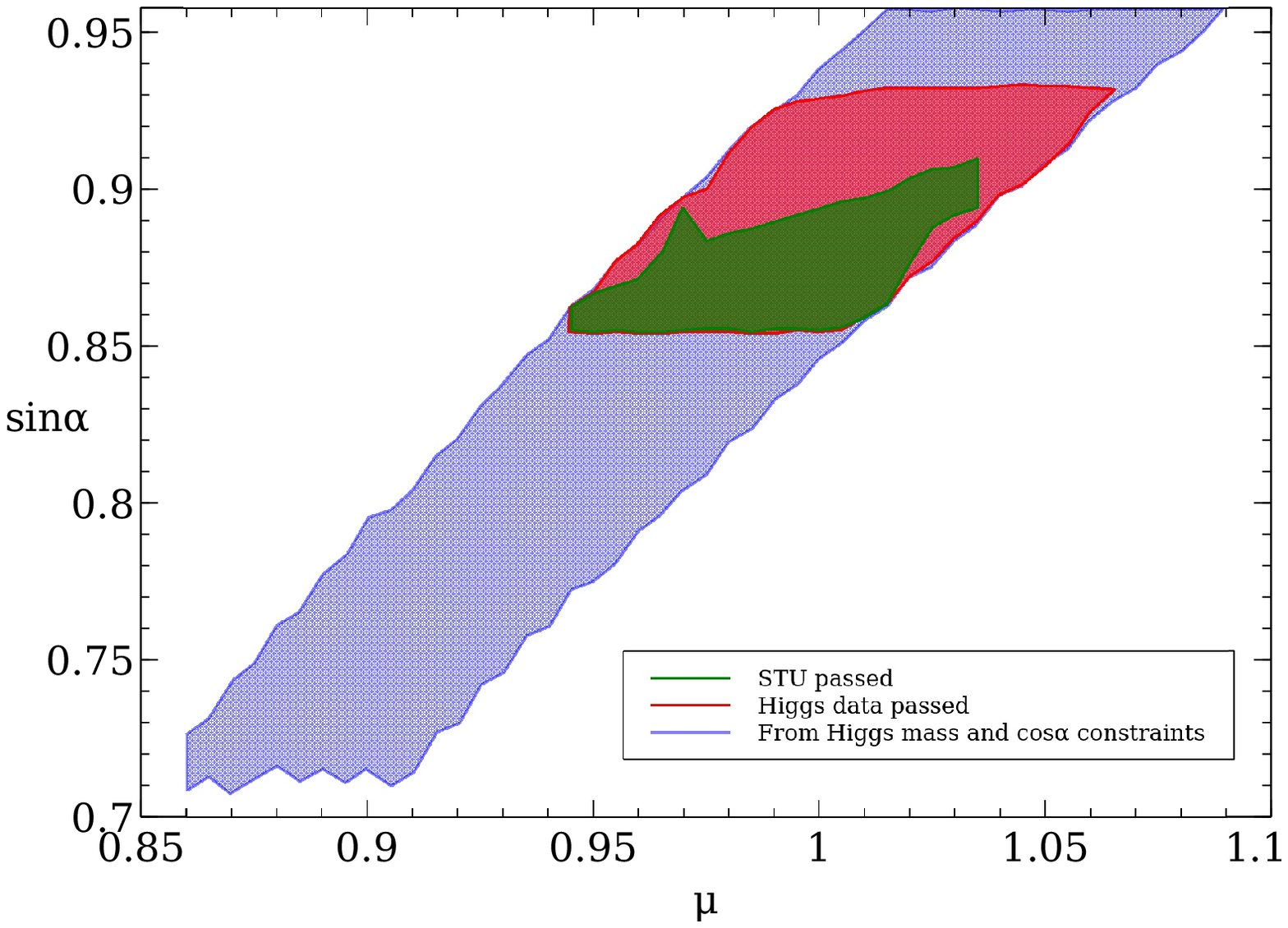}
%%%
\caption{\small Scatter plots for the allowed regions in the parameter space: $\sin\alpha$ as a function of $\lambda$ (left); $\sin\alpha$ as a function of $\mu$ (right). The blue shaded regions are allowed by the Higgs mass constraints, the red shaded regions are then allowed by the Higgs data ($3\sigma$) while the green shaded 
regions are then allowed by the S,T and U constraints ($1\sigma$). In both the plots, apart from the axes shown, all the 
other parameters were also  varied over their allowed ranges.}
\label{sin-alpha-lam-mu}
\end{figure}
%%%
parameters are also varied. Thus the projection of the allowed parameter space in the above three planes shows that $\lambda$ has an allowed range extending beyond $0.37$ and 
$\mu$ has an approximate allowed range of $0.94$-$1.06$, when we consider the Higgs data constraints to lie within 3$\sigma$ errors. 
We also find viable regions of the parameter space with $S,T$ and $U$ constraints now at only 1$\sigma$.

\begin{figure}[ht!]
\centering
\includegraphics[width=2.9in,height=2.3in]{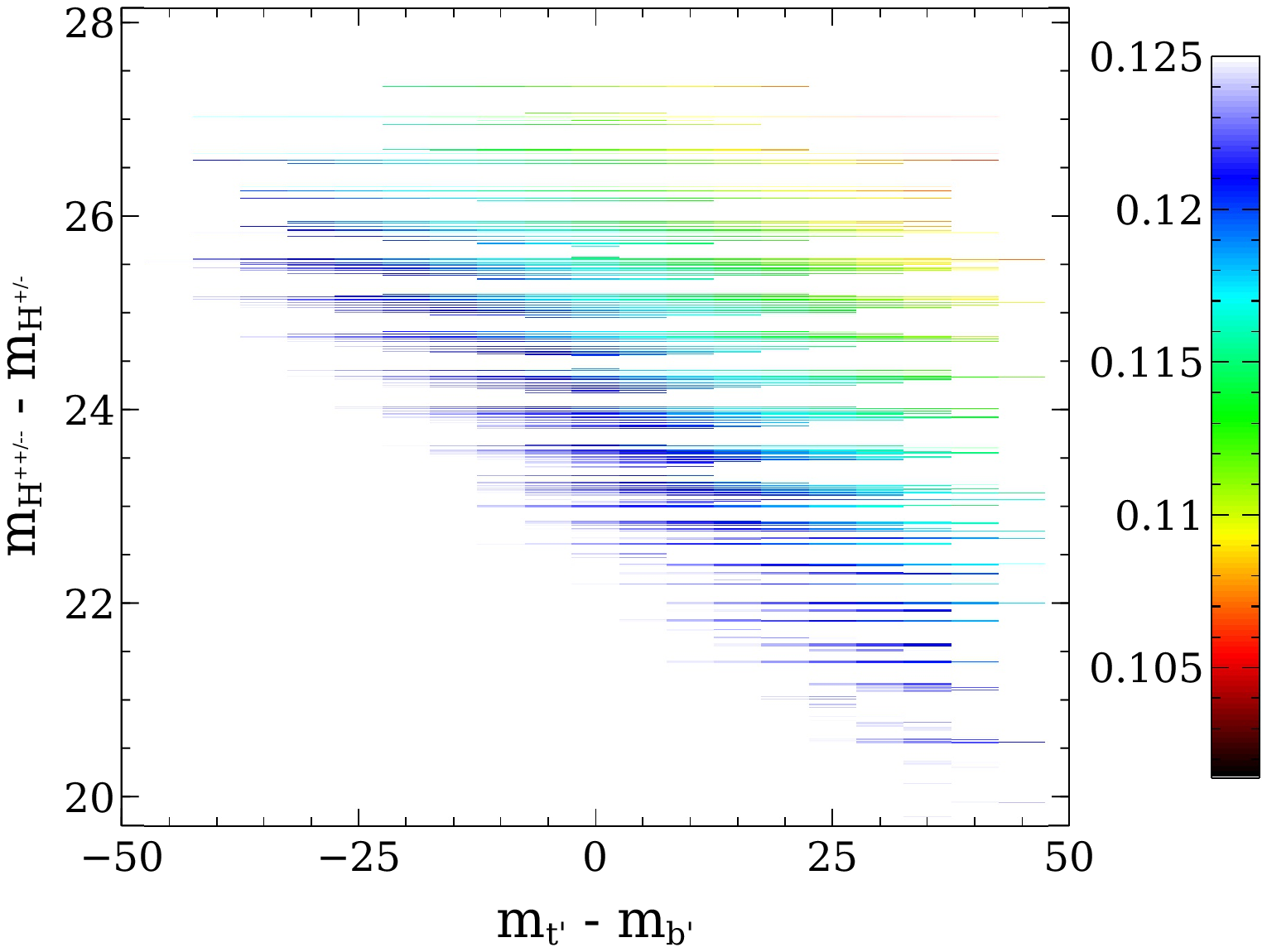}
\includegraphics[width=2.9in,height=2.3in]{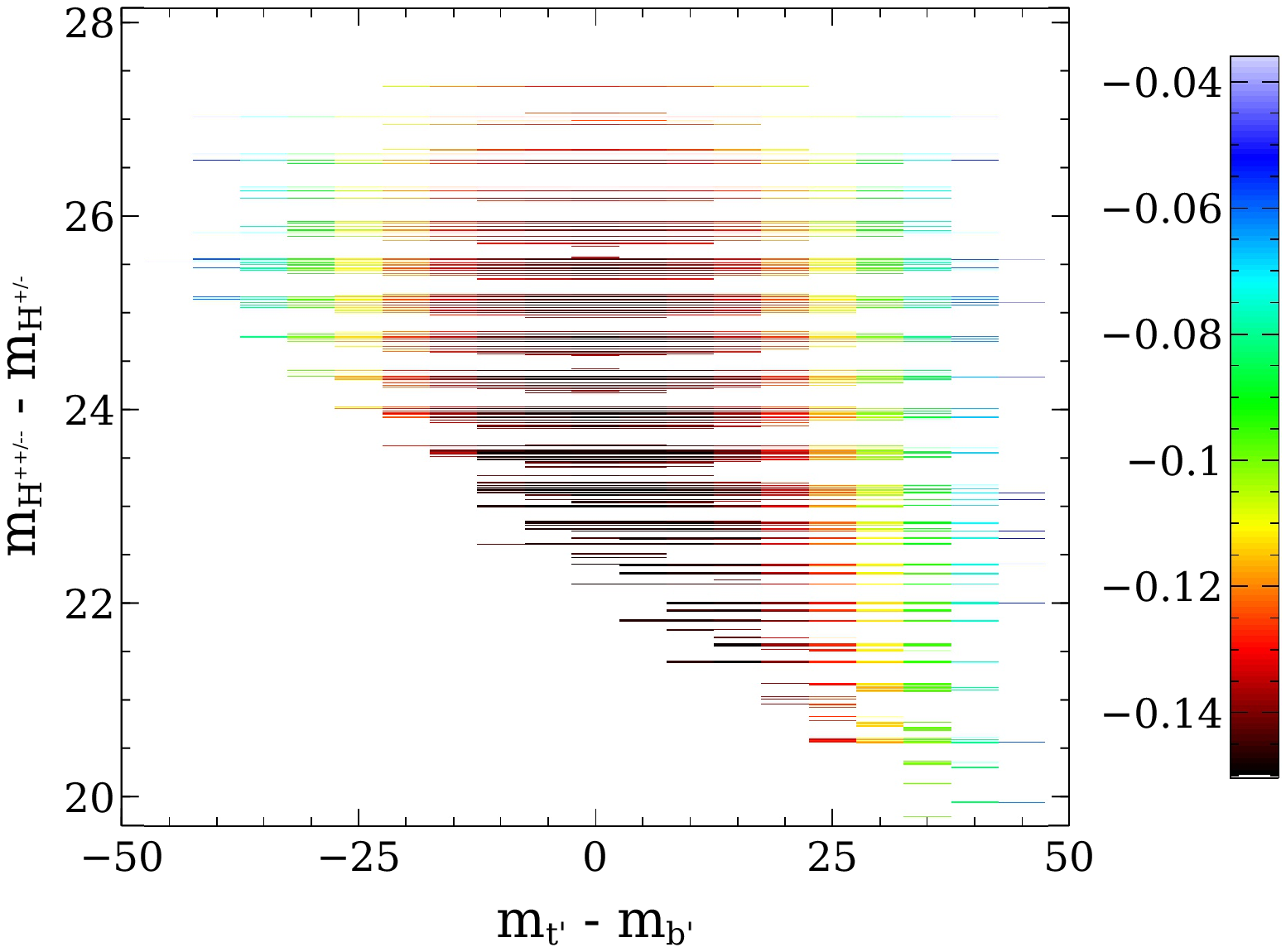}
\caption{\small Contour plots for $\Delta S$ (left) and $\Delta T$ (right) with $m_{t^{'}} - m_{b^{'}}$ and 
$m_{H^{\pm \pm}} - m_{H^{\pm}}$ varied. We vary $m_{t'}, m_{b'} \sim 550-600$ GeV which is consistent
with Higgs data within errors of 2$\sigma$. The color index at the right side of each plot shows the variation 
of $\Delta S$ (left) and $\Delta T$ (right). }
\label{delmtp-mbp-mHpp-mHp-delS-delT-1}
\end{figure}

\begin{figure}[ht!]
\centering
\includegraphics[width=3.1in,height=2.5in]{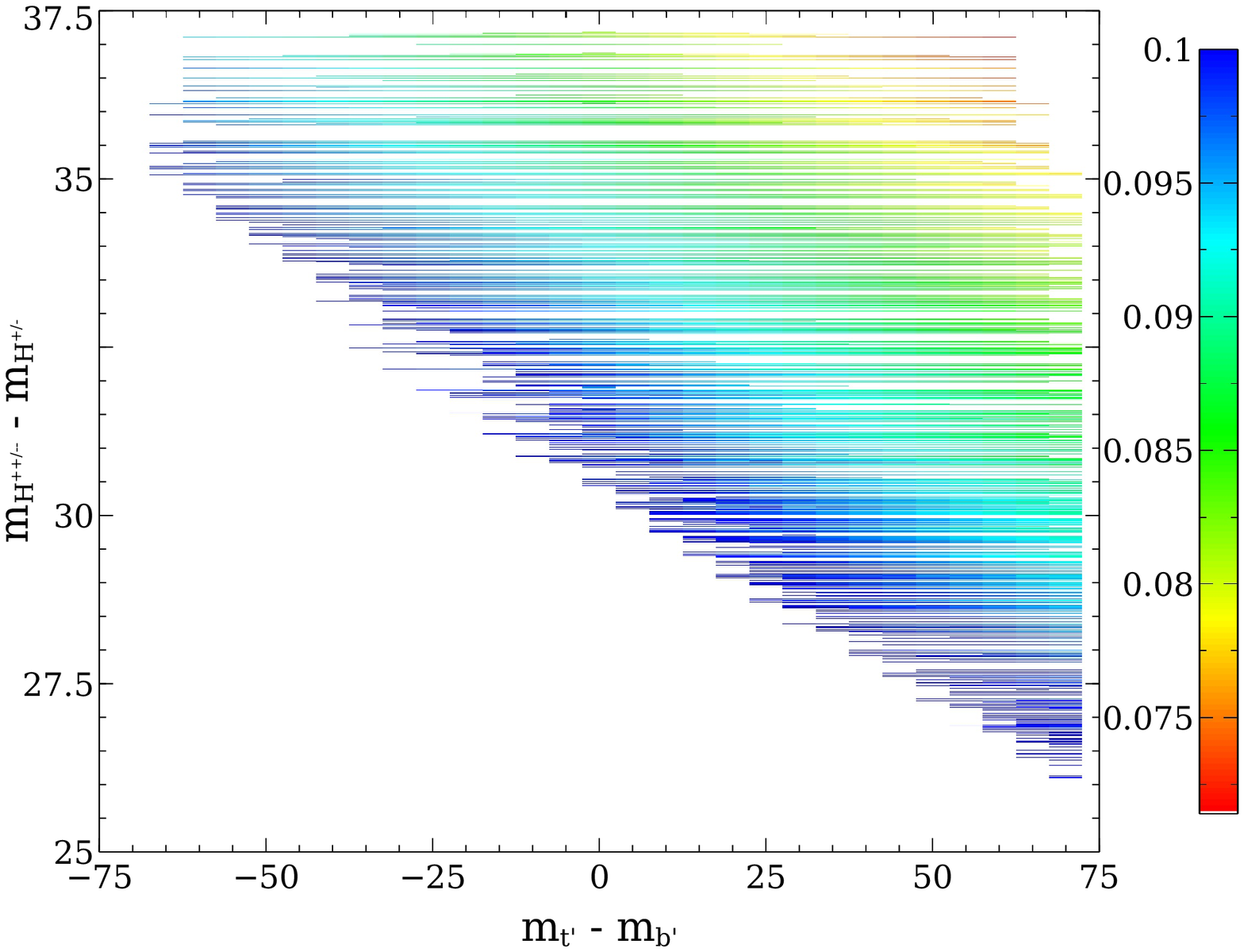}
\includegraphics[width=3.1in,height=2.5in]{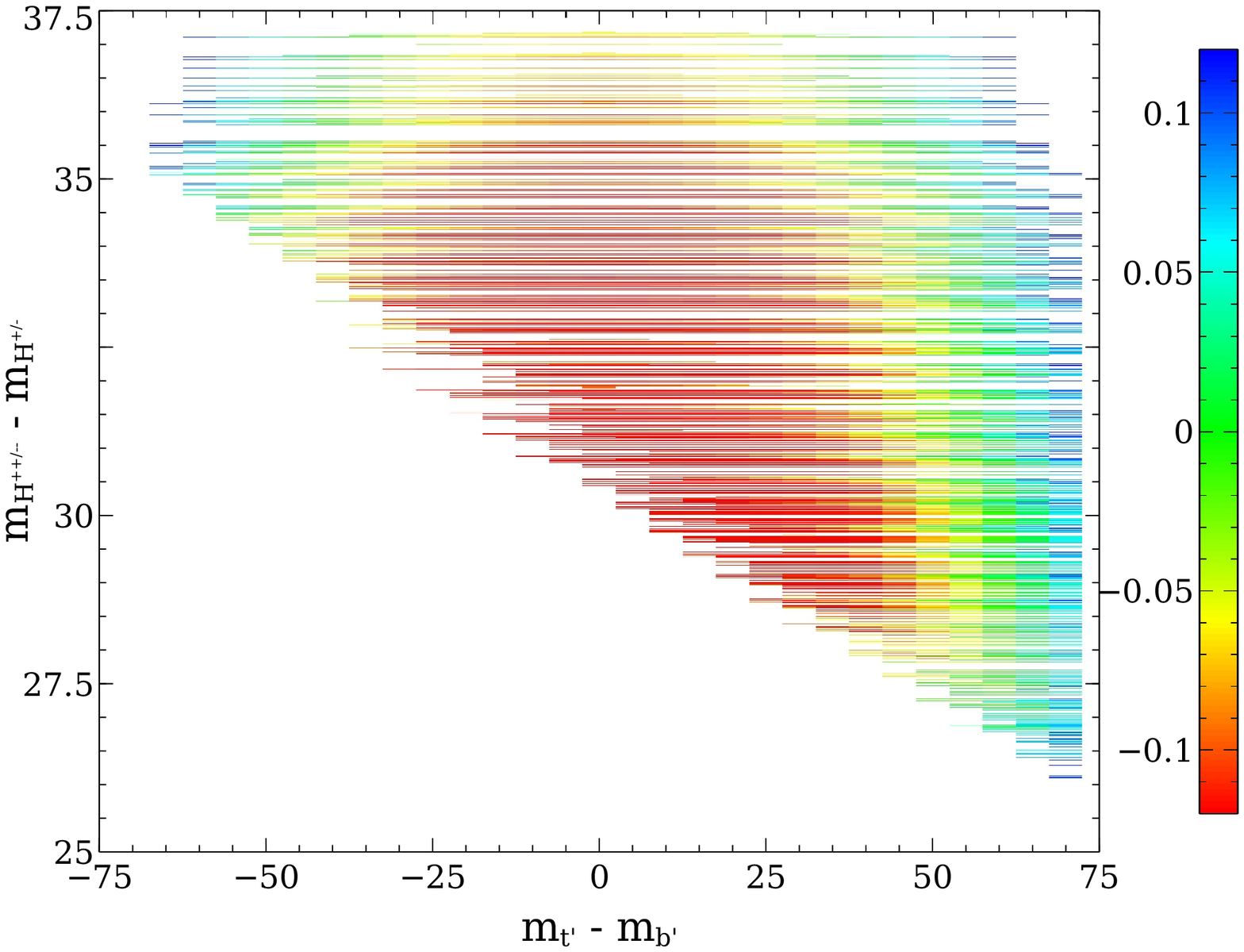}
\caption{\small Contour plots for $\Delta S$ (left) and $\Delta T$ (right) with $m_{t^{'}} - m_{b^{'}}$ and 
$m_{H^{\pm \pm}} - m_{H^{\pm}}$ varied. We vary $m_{t'}, m_{b'} \sim 550-750$ GeV which is consistent
with Higgs data within errors of 3$\sigma$. The color index at the right side of each plot shows the variation of $\Delta S$ (left) and $\Delta T$ (right). }
\label{delmtp-mbp-mHpp-mHp-delS-delT-2}
\end{figure}

As the survival of the fourth generation quarks in the current framework is closely related to the 
nontrivial mixing in the neutral scalar sector of our model, it becomes imperative to consider the 
allowed values of the mixing angle ($\alpha$) which also satisfies the various constraints.    
In Fig.~\ref{sin-alpha-lam-mu}, we show the allowed parameter space in the $\sin\alpha$-$\lambda$ and $\sin\alpha$-$\mu$ plane. Even though $\alpha$ is a function of $\lambda$ and $\mu$ (see Eq.~\ref{tan-beta-alpha}), these plots help us to emphasize  the 
importance of $\alpha$. As discussed before, we not only require $\alpha$ to suppress the hugely  
enhanced cross section of the Higgs production from gluon fusion, but $\alpha$ also determines the mass spectrum 
for the neutral scalars. Note that the scalar mass spectrum consistent with a $\sim 125$ GeV Higgs boson is allowed
over a range of $\sin\alpha$ lying between $\sim 0.7-0.95$. We find this a good place to discuss the features 
of the scalar mass spectrum that we obtain in our scan over the parameter space for $v_\Delta=3$ GeV. 
Note that we have already specified the requirement for the mass of the two CP-even neutral scalars to be $m_H \sim 97-99$ GeV and $m_h \sim 123-127$ GeV (Section \ref{sec:analysis}). The other scalars in the model are the pseudoscalar, singly and doubly charged scalars. For most of the allowed parameter values, 
the pseudoscalar has mass in the range $116 \leq m_A  \leq 123$ GeV as it has a linear dependence on the choice of $\mu$, once the VEVs are fixed. The choice of $\lambda_4$ already decides the mass hierarchy for the charged scalars. The singly charged scalars lie within the mass range of $\sim 118-214$ GeV while the doubly charged scalars are in the mass range of $\sim 125-276$ GeV when satisfying the Higgs mass constraints. We find that $\sin\alpha$ ranges roughly between $0.85$-$0.93$ (red region) 
solely from the Higgs data. This means that $\cos^{2}\alpha$ varies between $0.14$-$0.28$.  The allowed spectrum shrinks considerably once the oblique parameter corrections are included, yielding the 
modified ranges for 
the charged scalars as $m_{H^\pm} \sim 152-170$ GeV and $m_{H^{\pm\pm}} \sim 178-208$ GeV. This 
range features in the plots for mass splittings in Fig. \ref{delmtp-mbp-mHpp-mHp-delS-delT-1} and with 
slight variations (because of the weaker constraints being used from Higgs data) in Fig. \ref{delmtp-mbp-mHpp-mHp-delS-delT-2}. The other notable parameters that are worth mentioning are $\lambda_1$ which prefers
a large value to allow for large mixing in the neutral scalars, while values of $0 < \lambda_{2,3} < 5$ are preferred for the final allowed regions illustrated in our plots. It is worth pointing 
out that the triplet multiplet will
%%%
%%%
require a very weak Yukawa coupling to the lepton doublets (since $v_\Delta$ is large) to generate the tiny neutrino masses (see Eq. \ref{mn}), and therefore the doubly-charged scalars in the model will 
not have the usual 
dominant leptonic decay modes which has been extensively used by experimentalists to put limits on their mass, while the singly-charged Higgs bosons masses are within their experimental limits \cite{particledata}.
This ensures that the mass spectrum we obtain is safe from existing collider limits on such scalars. A detailed 
analysis of the collider signals for such scalars is left for future work.

Both the fourth generation fermions and the Higgs bosons in this model (HTM4)  significantly affect 
the oblique corrections through their contributions to the self energies of the gauge bosons. We highlight 
the mass splittings for the fourth generation quarks and the charged scalars in the model which give  
a parameter space that respects the Higgs data limits as well as the constraints from the oblique corrections.
Note that for our analysis, we have assumed degenerate fourth generation leptons ($m_{\tau'}=m_{\nu'}=250$ GeV).   
We illustrate our findings through the contour plots of $\Delta S$ and $\Delta T$ in the plane of $m_{t^{'}}-m_{b^{'}}$ 
versus $m_{H^{\pm \pm}}-m_{H^{\pm}}$ (see Fig.~\ref{delmtp-mbp-mHpp-mHp-delS-delT-1}
and  Fig.~\ref{delmtp-mbp-mHpp-mHp-delS-delT-2}). The colour coded index in each case shows 
the values of $\Delta S$ (for the left curve) and $\Delta T$ (for the right curve). The results are plotted for  
$m_{t'}, m_{b'} \sim 550-600$ GeV in Fig.~\ref{delmtp-mbp-mHpp-mHp-delS-delT-1}, and for  $m_{t'}, m_{b'} \sim 550-750$ GeV in Fig.~\ref{delmtp-mbp-mHpp-mHp-delS-delT-2}. While in Figs.~\ref{lam4-mu-lam1}  and  \ref{lam4-mu-lam2}, the parameter space varied only slightly when 
increasing the fourth generation masses, the 
%%%
change is much more noticeable in here, where a larger region of the parameter 
space survives for $m_{t'}, m_{b'} \sim 600-750$ GeV.
An interesting feature to note is that even though the contributions in HTM4 for 
$\Delta S$ is found to be only positive, the contributions of the model in $\Delta T$ is found to be 
both positive and negative. The negative contributions to the $\Delta T$ come from the bosonic loops and from 
the extended scalar sector, with the non trivial mixing also playing a significant role. From the plots, we find 
that $m_{t^{'}}$ can be either greater than or less than $m_{b^{'}}$ for the allowed parameter space. On the other hand, $m_{H^{\pm \pm}}$ is always greater than the $m_{H^{\pm}}$, due to our original 
choice requiring  $\lambda_{4}$ to be negative. 
Clearly we find that a larger mass splitting is allowed in HTM4 when compared to the SM4 which is disfavoured from Higgs data alone. 
 
Note that the mixing angle $\alpha$ will suppress the production cross section of the $\sim 125$ GeV Higgs boson in the $VBF$ and $VH$ channels. Although, these channels, when considered inclusively along with the gluon fusion channel are subdominant, we check the rates for them against the 
values shown in Table~\ref{tab:tab1} for a representative 
point which has passed all the constraints discussed earlier. For the point $\sin\alpha \approx 0.87$, $\lambda = 0.365$, $\mu = 1.0$, $\lambda_{4} = -0.79$, $m_{t'} = 600$ GeV, $m_{b'} = 650$ GeV in the parameter space, 
we find that $\frac{[\sigma \times BR]^{SM4+triplet}_{VBF}}{[\sigma \times BR]^{SM}_{VBF}}\simeq 0.062$ in the $WW^{*}$ channel ($2.11\sigma$ away 
from the combined central value of ATLAS and CMS) and $\frac{[\sigma \times BR]^{SM4+triplet}_{VH}}{[\sigma \times BR]^{SM}_{VH}}\simeq 0.243$. in the $b\bar{b}$ channel ($1.16\sigma$ away from the combined central value of ATLAS and CMS). In fact we find that for all the STU passed points, the 
$VBF$ signal strengths for the $WW^{*}$ channel are within $\sim 2.15\sigma$  while the $VH$ signal strengths for the $b\bar{b}$ channel are within 
$\sim 1.35\sigma$ from their respective experimental signal strengths.

\section{Summary and Conclusion}
\label{sec:conclusions}
An extension of the SM by a fourth sequential generation of quarks and leptons is one of the simplest beyond the SM scenarios. The fourth generation rectifies some of the shortcomings of the SM, and in particular, it provides a dark matter candidate, $\nu^\prime$. Electroweak precision measurements (oblique and non-oblique corrections) favour small mass splittings between the fourth generation quarks, $m_{t^\prime}-m_{b^\prime}< M_W$, and flavor violating decays are restricted by the structure of the $3 \times 3$ CKM matrix.

However, for all its nice features, the fourth generation has all but been abandoned. This is mainly due to two experimental inputs. First, discovery of the Higgs boson and recent measurements of the Higgs production cross section and decay rates disfavor the SM with four generations. The Higgs production through gluon fusion is enhanced by the presence of new, $SU(3)_c$ interacting heavy particles, rendering it inconsistent with experimental results associated with the Higgs boson at $\sim 125$ GeV. Second, direct limits on the masses of fourth generation quarks, all assuming that the decays  $t^\prime \to Wb$   and $b^\prime \to W t$ have 100\% branching ratios, have pushed limits on the fourth generation quark masses into the 700-800 GeV region. This has serious implications on Yukawa couplings, with perturbativity of the couplings in serious jeopardy. 

In this work we show that, relaxing the single Higgs boson requirement and  slightly relaxing the experimental bounds (that is, allowing for alternative decays of the fourth generation fermions) saves the fate of the fourth generation. We choose to work in the Higgs Triplet Model, where the neutral triplet Higgs state is allowed to mix non trivially with the SM doublet Higgs state. Unlike in Two Higgs Doublet models, in this model the Yukawa couplings are not proportional to the ratio of the doublet VEVs, lifting some of the pressure on the perturbativity of couplings. We allow for a second neutral Higgs boson, just below the sensitivity of  both Tevatron and LHC, but consistent with the $2\sigma$ bump at LEP in the 98-99 GeV region. We impose precision electroweak constraints on the model, constraints on the masses of the singly and doubly charged Higgs masses, as well as requirement that the production and decay rates we obtain reproduce the data from ATLAS and CMS, within  their errors. We scan the parameter space of the HTM4 model under these restricted conditions and show that some regions of the parameter space, {\it albeit} restricted but still significant, survives. Couplings in the Higgs potential are limited to small regions: $\lambda$ has an allowed range between  $0.34$-$0.375$,  $\mu$ has an approximate allowed range between  $0.94$-$1.07$, and $\lambda_{4}$, chosen to be negative, 
has an allowed range between $-0.9$ and $-0.1$. The precise value for $\sin\alpha$, which ranges roughly  between $0.85$-$0.93$ from the Higgs data at $3\sigma$, is consistent with the other constraints. We find that $m_{t'}$ can be either larger or smaller than $m_{b'}$, and the parameter space for $m_{b'}, m_{t'}$ in the 550-600 GeV region is under less pressure than the one for $m_{b'}, m_{t'}$ in the 600-750 GeV region. Our analysis clearly shows that, when even after imposing all collider and electroweak precision constraints, some regions of the parameter space survive, giving support to the hypothesis that the fourth generation is ruled out in models with a single doublet Higgs only, and survives when one adds a triplet Higgs field.  We also find that a wide range of mass values for the sequential fourth generation quarks ($\sim 550-750$ GeV) is allowed by data provided a very light (single- and double-) charged scalar spectrum with mass less than $210$ GeV exists. These mass ranges are well within the reach of the current LHC experiment and would be a perfect testing ground to search for the complementary signals of the scalars and fermions of HTM4 in the current data and future runs.     

\section{Acknowledgement}

SB would like to thank Titas Chanda and Animesh Chatterjee for some technical help. The work of SB and SKR was partially supported by funding available from the Department of Atomic Energy, 
Government of India, for the Regional Centre for Accelerator-based Particle Physics, Harish-Chandra 
Research Institute. The work of  M.F.  is supported in part by NSERC under grant number
SAP105354.

\section{Appendix}
\label{appendix}
\subsection{Higgs Decays in HTM4}
\label{appendix1}
In the HTM, in addition to the SM-like Higgs boson $h$, 
there are  doubly-charged scalar bosons $H^{\pm\pm}$, singly-charged scalar bosons $H^\pm$,  one 
neutral CP-even  scalar boson  $H$ and one CP-odd scalar $A$. 
Their decays are affected by the presence of the fourth generation. Below we present the decays for the neutral CP-even Higgs bosons $h$ and $H$, as these are most likely to be candidates for the state at $\sim 125$ GeV observed at the LHC.

\subsubsection{\bf  Decay rates of $h$}
\label{subsec:decay}
In the limit $v_\Delta \to 0$ the triplet Higgs decouples from the potential, and $h$ is the SM-like Higgs boson.

\bigskip

$\bullet$~~~~{\bf Tree-level decay rates for $h\to f \bar f,~WW, ~WW^*,~ZZ,~ZZ^*$}

The decay rates for $h$ can be expressed as 
\begin{eqnarray}
\Gamma(h\to f\bar{f})&=&G_F\frac{m_hm_f^2}{4\sqrt{2}\pi}N_c^f\beta\left(\frac{m_f^2}{m_h^2}\right)^3\cos^2\alpha,\\
\Gamma(h \to \nu \nu)&=&\Gamma(h \to \nu^c \bar{\nu})+\Gamma(h \to \bar{\nu}^c \nu)
=\sum_{i,j=1}^4S_{ij}|h_{ij}|^2\frac{m_h}{4\pi}\sin^2\alpha,
\end{eqnarray}
where $\beta(x)=\sqrt{1-4x}$. 
The decay rate of the Higgs boson $h$ decaying into the gauge boson  pair $VV$ ($V=W$ or $Z$) is given by 
\begin{eqnarray}
\Gamma(h\to VV)&=&\frac{ |\kappa_V(h)|^2m_h^3}{128 \pi m_V^4}    \delta_V\left[1-\frac{4m_V^2}{m_h^2}+\frac{12m_V^4}{m_h^4}\right]\beta\left(\frac{m_V^2}{m_h^2}\right)
\end{eqnarray}
where $\delta_W=2$ and $\delta_Z=1$, and where $\kappa_V(h)$ are the couplings of the Higgs $h$ with the vector bosons:
\begin{eqnarray}
\kappa_W(h)&=&\frac{ig^2}{2} \left( v_\phi\cos \alpha+ 2 v_\Delta \sin \alpha \right), \\
\kappa_Z(h)&=&\frac{ig^2}{2 \cos^2\theta_W} \left( v_\phi\cos \alpha+ 4 v_\Delta \sin \alpha \right)
\end{eqnarray}
The decay rates for the three body decay modes are, 
\begin{eqnarray}
\Gamma(h\to VV^*)&=&\frac{3g_V^2 |\kappa_V(h)|^2m_h}{512 \pi^3 m_V^2}\delta_{V'}F(\frac{m_V^2}{m_h^2}),\end{eqnarray}
where $\delta_{W'}=1$ and $\displaystyle \delta_{Z'}=\frac{7}{12}-\frac{10}{9} \sin^2 \theta_W +\frac{40}{27} \sin^4 \theta_W$, and  
the function $F(x)$ is given as 
\begin{eqnarray}
F(x)&=&-|1-x|\left(\frac{47}{2}x-\frac{13}{2}+\frac{1}{x}\right)+3(1-6x+4x^2)|\log \sqrt{x}|\nonumber\\
&+&\frac{3(1-8x+20x^2)}{\sqrt{4x-1}}\arccos \left(\frac{3x-1}{2x^{3/2}}\right).
\label{decay_F}
\end{eqnarray}

Explicitly, inserting the couplings for the HTM4:
\begin{eqnarray}
\Gamma(h\to W^+W^-)&=&\frac{g^4 m_h^3}{64\pi m_W^4}\left({v_\phi}\cos\alpha +2v_\Delta \sin\alpha\right)^2\left(\frac{1}{4}-\frac{m_W^2}{m_h^2}+\frac{3m_W^4}{m_h^4}\right)\beta\left(\frac{m_W^2}{m_h^2}\right),~~~~~~\\
\Gamma(h\to ZZ)&=&\frac{g_Z^4 m_h^3}{128\pi m_Z^4}\left({v_\phi}\cos\alpha +4v_\Delta \sin\alpha\right)^2\left(\frac{1}{4}-\frac{m_Z^2}{m_h^2}+\frac{3m_Z^4}{m_h^4}\right)\beta\left(\frac{m_Z^2}{m_h^2}\right),\\
\Gamma(h\to WW^*)&=&\frac{3 g^6 m_h}{2048\pi^3m_W^2}(v_\phi\cos\alpha+2v_\Delta \sin\alpha)^2 F\left(\frac{m_W^2}{m_h^2}\right),\\
\Gamma(h\to ZZ^*)&=&\frac{g_Z^6 m_h}{8192\pi^3 m_Z^2}({v_\phi}\cos\alpha+4v_\Delta \sin\alpha)^2\nonumber\\
&&\times\left(7-\frac{40}{3}\sin^2\theta_W+\frac{160}{9}\sin^4\theta_W\right)F\left(\frac{m_Z^2}{m_h^2}\right),
\end{eqnarray}

\bigskip

$\bullet$~~~{ \bf Loop-induced decay rates for $h\to \gamma\gamma,~gg$}
\begin{eqnarray}
\label{eq:DTHM-h2gaga}
\Gamma(h \rightarrow\gamma\gamma)
& = & \frac{G_F\alpha^2 m_{h}^3}
{128\sqrt{2}\pi^3} \bigg| \sum_f N^f_c Q_f^2 g_{h ff} 
A_{1/2}^{h}
(\tau_f) + g_{h WW} A_1^{h} (\tau_W) \nonumber \\
&& + {g}_{h H^\pm\,H^\mp}
A_0^{h}(\tau_{H^{\pm}})+
 4 {g}_{h H^{\pm\pm}H^{\mp\mp}}
A_0^{h}(\tau_{H^{\pm\pm}}) \bigg|^2 \, .
\label{partial_width_htm}
\end{eqnarray}
Here $\alpha$ is the fine-structure constant, $N_c(=3)$ is the number of quark colors, $Q_f$ is the
electric charge of the fermion in the loop, and $\tau_{i}=m^2_{h}/4m^2_{i}$ $(i=f,W,H^{\pm},H^{\pm\pm})$. 
The loop functions $A_1$ (for the $W$ boson) and $A_{1/2}$
(for the fermions, $f$) are  
\begin{eqnarray}
A_{1/2}^h(\tau)&=&2 \left[ \tau+(\tau-1)f(\tau)\right] \tau^{-2},
\label{eq:Afermion} \\
A_1^h(\tau)&=& -\left [2 \tau^2+3 \tau +3(2 \tau -1)f(\tau) \right ] \tau^{-2},
\label{eq:Avector} 
\end{eqnarray}
For the contribution from the fermion loops we will only keep the term with the $t,~b'$ and $t'$ quarks,
which are dominant. The loop function for $H^{\pm\pm}$ and $H^{\pm}$ is given by:
\begin{eqnarray}
A_{0}^{h}(\tau) &=& -[\tau -f(\tau)]\, \tau^{-2} \, ,
\label{eq:Ascalar}
\end{eqnarray}
and the function $f(\tau)$ is given by
\begin{eqnarray}
f(\tau)=\left\{
\begin{array}{ll}  \displaystyle
\arcsin^2\sqrt{\tau} & \tau\leq 1 \\
\displaystyle -\frac{1}{4}\left[ \log\frac{1+\sqrt{1-\tau^{-1}}}
{1-\sqrt{1-\tau^{-1}}}-i\pi \right]^2 \hspace{0.5cm} & \tau>1 \, .
\end{array} \right. 
\label{eq:ftau} 
\end{eqnarray}
Note that the contribution from the loop with $H^{\pm\pm}$ in Eq. \ref{partial_width_htm} 
is enhanced relative to the contribution from $H^{\pm}$ by a factor of four at the amplitude level.
The couplings of $h$ to the vector bosons and fermions relative to the values in the SM are as follows:
\begin{eqnarray}
g_{h f \bar f}\equiv g_{h t\overline t, h b'\overline {b'}, h t'\overline {t'}}=\cos\alpha/\cos\beta_{\pm} \, ,
\label{h1tt}\\
g_{hWW}= \cos\alpha+2\sin\alpha v_\Delta/v_\Phi  \, ,  
\label{h1WW} \\
g_{hZZ}= \cos\alpha+4\sin\alpha v_\Delta/v_\Phi\ \, .
\label{h1ZZ}
\end{eqnarray}

The scalar trilinear couplings are parametrized as follows \cite{Arhrib:2011uy}:
\begin{eqnarray}
{g}_{h H^{++}H^{--}}  & = & \frac{m_W}{ g m_{H^{\pm \pm}}^2}  \{2\lambda_2v_\Delta \sin\alpha+\lambda_1v_\Phi \cos \alpha\} \,
 \label{eq:redgcalHHpp}\\
{g}_{h H^+H^-} & = &  \frac{m_W}{2g m_{H^{\pm}}^2} 
\bigg\{\left[4v_\Delta(\lambda_2 + \lambda_3) \cos^2{\beta_{\pm}}+2v_\Delta\lambda_1\sin^2{\beta_{\pm}}-
\sqrt{2}\lambda_4v_\Phi \cos{\beta_{\pm}} \sin{\beta_{\pm}}\right ]\sin\alpha  \nonumber \\ 
&+&\left [\lambda\,v_\Phi \sin{\beta_{\pm}}^2+{(2\lambda_{1}+\lambda_{4}) }v_\Phi \cos^2{\beta_{\pm}}+
(4\mu-\sqrt{2}\lambda_4v_\Delta)\cos{\beta_{\pm}}\sin{\beta_{\pm}}\right ] \cos\alpha\bigg\}.~~~~~
\label{eq:redgcalHHp}
\end{eqnarray}

The loop induced decay process $h\to gg$  can be  expressed by
\begin{eqnarray}
\Gamma(h\to gg)&=&\frac{G_F\alpha_s^2m_h^3}{64\sqrt{2}\pi^3}\cos^2\alpha |\sum_{q=1}^4A_{1/2}^h (\tau_q)|^2,
\end{eqnarray}
 where the loop functions are given by Eq. \ref{eq:Afermion},  
with $f(\tau)$ given by Eq. \ref{eq:ftau}.

Finally, the loop induced decay process $h \to Z \gamma$ is 
\begin{eqnarray}
\Gamma(h\to Z\gamma)&=&\frac{G_F\alpha_{\rm{em}}^2m_h^3}{64\pi^4}\left(1-\frac{m_Z^2}{m_h^2}\right)^3
\Big |\sum_f N_f^c Q_fg_{hf\bar f}A^h_{1/2}(\tau^{-1}_f, \lambda^{-1}_f)+g_{hWW} A_1^h(\tau^{-1}_W, \lambda^{-1}_W) \nonumber \\
&+& (g_{hH^+ H^-})(g_{ZH^+H^-})A_0^h(\tau^{-1}_{H^+}, \lambda^{-1}_{H^+})+ (g_{hH^{++} H^{--}})(g_{ZH^{++}H^{--}})A_0^h(\tau^{-1}_{H^{++}}, \lambda^{-1}_{H^{++}})\Big |^2, \nonumber\\
\end{eqnarray}
with $\displaystyle \tau_i=\frac{m_h^2}{4m_i^2}, ~\lambda_i=\frac{m_Z^2}{4m_i^2}$, and where the loop functions are given by
\begin{eqnarray}
A^h_{1/2}(\tau^{-1}, \lambda^{-1})&=&\frac{2}{\cos\theta_W}(I_{3f}-2Q_f\sin^2\theta_W) \left[ I_1(\tau^{-1}, \lambda^{-1})-I_2(\tau^{-1}, \lambda^{-1}) \right], \label{eq:AfermionZ} \qquad \\
A_1^h(\tau^{-1}, \lambda^{-1})&=&\cos \theta_W  \left\{4\left(3-\tan^2\theta_W\right)I_2(\tau^{-1}, \lambda^{-1}) \right. \nonumber\\
&+& \left. \left[\left(1+{2}{\tau}\right)\tan^2\theta_W-\left(5+{2}{\tau}\right)\right]I_1(\tau^{-1}, \lambda^{-1}) \right\}, \label{eq:AvectorZ} \\
A_0^h (\tau^{-1}_{H^c}, \lambda^{-1}_{H^c})&=&I_1(\tau^{-1}_{H^c}, \lambda^{-1}_{H^c}), \quad c=+, ++\label{eq:AscalarZ} \\
I_1(\tau^{-1}, \lambda^{-1})&=&-\frac{1}{2(\tau-\lambda)}+\frac{1}{2 (\tau-\lambda)^2}\left[f(\tau)-f(\lambda)\right] \nonumber \\
&+&\frac{ \lambda}{(\tau-\lambda)^2}\left [g(\tau)-g(\lambda) \right] \\
I_2(\tau^{-1}, \lambda^{-1})&=&\frac{1}{2 (\tau-\lambda)}\left[f(\tau)-f(\lambda)\right]  
\label{eq:loopf}
\end{eqnarray}
In the equations above, 
\begin{eqnarray}
f(\tau)=\Bigg\{
\begin{array}{c}
\left [\arcsin(\sqrt{\tau}) \right ]^2, \quad \rm{if }~\tau\leq 1,\\
-\displaystyle\frac{1}{4}\left [\ln  \frac{1+\sqrt{1-\tau^{-1}}}{1-\sqrt{1-\tau^{-1}}}-i\pi\right ]^2, \quad \rm{if }~\tau> 1
\end{array}. 
\label{eq:ftau}
\end{eqnarray}
and 
\begin{eqnarray}
g(\tau^{-1})=\Bigg\{
\begin{array}{c}
\sqrt{\tau-1} \left [\arcsin(\sqrt{\frac{1}{\tau}}) \right ], \quad \rm{if }~\tau > 1,\\
\displaystyle\frac{\sqrt{1-\tau}}{2}\left [\ln \displaystyle \frac{1+\sqrt{1-\tau}}{1-\sqrt{1-\tau}}-i\pi\right ], \quad \rm{if }~\tau \leq 1
\end{array}. 
\label{eq:gtau}
\end{eqnarray}
and the couplings between the charged Higgs bosons and the $Z$ are:
\begin{eqnarray}
g_{ZH^+H^-}&=&\frac{2}{ \sin 2\theta_W} \frac{v_\Delta^2 \cos 2\theta_W- v_\phi^2 \sin \theta_W^2}{v_\phi^2+ 2 v_\Delta^2} \\
g_{ZH^{++}H^{--}}&=&\frac{2\cos 2 \theta_W}{\sin 2 \theta_W}
\end{eqnarray}

\subsubsection{\bf  Decay rates of $H$}
In the limit in which $\alpha \to 0$, the $H$ Higgs boson is mostly triplet-like, thus this is the non SM-like boson. The decays of $H$ can be obtained from the corresponding formulas for $h$, with the substitution $ \cos \alpha \to - \sin \alpha, \, \sin \alpha \to \cos \alpha$. 

\subsection{Constraints on the Higgs Potential}
\label{appendix2}
These have been thoroughly analyzed in \cite{Arhrib:2011uy}, and we summarize their results briefly. Positivity requirement in the singly and doubly
charged Higgs mass sectors, and choosing $v_\Delta>0$,  require:
\begin{eqnarray}
\mu&>&0  \label{eq:tachyonodd} \\
\mu &>& \frac{\lambda_4 v_\Delta}{2\sqrt{2}} \label{eq:tachyon1} \\
\mu &>& \frac{\lambda_4 v_\Delta}{\sqrt{2}} + \sqrt{2} \frac{ \lambda_3 v_\Delta^3}{v_\Phi^2} \label{eq:cons1}
\end{eqnarray}
 while for the requirement that the potential is bounded from below the complete set of conditions are:
 \begin{eqnarray}
&& \lambda > 0 \;\;{\rm \&}\;\; \lambda_2+\lambda_3 > 0  \;\;{\rm \&}  \;\;\lambda_2+\frac{\lambda_3}{2} > 0 
\label{eq:potential1} \\
&& {\rm \&} \;\;\lambda_1+ \sqrt{\lambda(\lambda_2+\lambda_3)} > 0 \;\;{\rm \&}\;\;
\lambda_1+ \sqrt{\lambda(\lambda_2+\frac{\lambda_3}{2})} > 0  \label{eq:potential2}\\
&& {\rm \&} \;\; \lambda_1+\lambda_4+\sqrt{\lambda(\lambda_2+\lambda_3)} > 0 \;\; {\rm \&} \;\; 
\lambda_1+\lambda_4+\sqrt{\lambda(\lambda_2+ \frac{\lambda_3}{2})} > 0  \label{eq:potential3}
\end{eqnarray}
Of these, the last expression in Eq. (\ref{eq:potential3}) would restrict possible enhancements in the $h,\,H \to \gamma \gamma$ decay.

\end{document}